\documentclass[preprint,12pt]{aastex}  %--original for ApJ
\usepackage{epsfig}

\def\Hbs{H$\beta$~}
\def\oiii{O\,{\sc iii]}}
\def\oviii{O\,{\sc viii]}}
\def\ovii{O\,{\sc vii]}}

\def\feii{Fe\,{\sc ii}}
\def\feiii{Fe\,{\sc iii}}
\def\fevi{Fe\,{\sc vi}}
\def\fex{Fe\,{\sc x}}

\def\fexx{Fe\,{\sc xx}}
\def\fexiii{Fe\,{\sc xiii}}
\def\fexvii{Fe\,{\sc xvii}}
\def\fexxv{Fe\,{\sc xxv}}
\def\fexxvi{Fe\,{\sc xxvi}}

\def\ovii{O\,{\sc vii}}
\def\oviii{O\,{\sc viii}}

\def\cm{\ifmmode {\rm cm}^{-1} \else cm$^{-1}$ \fi}
\def\s{\ifmmode {\rm s}^{-1} \else s$^{-1}$ \fi}
\def\cc{\ifmmode {\rm cm}^{-3} \else cm$^{-3}$ \fi}
\def\cs{\ifmmode {\rm cm}^{-2} \else cm$^{-2}$ \fi}
\def\g{\ifmmode \gamma \else $\gamma$\fi}
\def\G{\ifmmode \Gamma \else $\Gamma$\fi}
\def\Gs{\ifmmode \Gamma~ \else $\Gamma~$\fi}

\def\gc{\ifmmode \gamma_{\rm c} \else $\gamma_{\rm c}$ \fi}
\def\sw{Schwarzschild~}
\def\gsim{\mathrel{\raise.5ex\hbox{$>$}\mkern-14mu
             \lower0.6ex\hbox{$\sim$}}}
\def\lsim{\mathrel{\raise.3ex\hbox{$<$}\mkern-14mu
             \lower0.6ex\hbox{$\sim$}}}
\def\simless{\mathbin{\lower 3pt\hbox
     {$\rlap{\raise 5pt\hbox{$\char'074$}}\mathchar"7218$}}}   %< or of order
\def\simmore{\mathbin{\lower 3pt\hbox
     {$\rlap{\raise 5pt\hbox{$\char'076$}}\mathchar"7218$}}}   %> or of order
\def\Msun{M_\odot}                                % solar masses
\def\4u{4U 1728--34}
\def\deg{^\circ}
\newcommand{\Alfven}{Alfv$\acute{\rm e}$n~}

\lefthead{et al.} \righthead{\4u}

\shorttitle{X-ray Absorbing MHD Winds from Accretion Disks}

\shortauthors{Fukumura, Kazanas, Contopoulos, Behar}

\begin{document}

%\title{MHD Accretion Disk Winds as AGN Warm Absorbers}
\title{MHD Accretion-Disk Winds as X-ray Absorbers in AGNs}

%\date{\today}

%
\author{\textsc{Keigo Fukumura}\altaffilmark{1,2,3},
\textsc{Demosthenes Kazanas}\altaffilmark{3},
\textsc{Ioannis Contopoulos}\altaffilmark{4}, \\
\textsc{and} \\
\textsc{Ehud Behar}\altaffilmark{3,5,6} }

\altaffiltext{1}{Email: Keigo.Fukumura@nasa.gov}

\altaffiltext{2}{University of Maryland, Baltimore County
(UMBC/CRESST), Baltimore, MD 21250}

\altaffiltext{3}{Astrophysics Science Division, NASA/Goddard Space
Flight Center, Greenbelt, MD 20771} \altaffiltext{4}{Research Center
for Astronomy, Academy of Athens, Athens 11527, Greece}
\altaffiltext{5}{Department of Physics, Technion, Haifa 32000,
Israel} \altaffiltext{6}{Senior NPP Fellow}

\begin{abstract}

\baselineskip=15pt

We present the two-dimensional (2D) ionization structure of
self-similar magnetohydrodynamic (MHD) winds off accretion disks
around irradiated by a central X-ray point source. Based
on earlier observational clues and theoretical arguments, we focus
our attention on a subset of these winds, namely those with radial
density dependence $n(r)\propto 1/r$ ($r$ is the spherical radial
coordinate).
%Assuming a power-law spectrum for the ionizing source
%, i.e. $F_{\nu} \propto \nu^{-\alpha} ~(\alpha \simeq 1.5)$,
We employ the photoionization code \verb"XSTAR" to compute the ionic
abundances of a large number of ions of different elements and then
compile their line-of-sight (LOS) absorption columns. We focus our
attention on the distribution of the column density of the various
ions as a function of the ionization parameter $\xi$ (or
equivalently $r$) and the angle $\theta$. Particular attention is
paid to the absorption measure distribution (AMD), namely their
Hydrogen-equivalent column per logarithmic $\xi$ interval, $d N_H/d
\log \xi$, which provides a measure of the winds' radial density
profiles. For the chosen density profile $n(r)\propto
1/r$ the AMD is found to be independent of $\xi$, in good agreement
with its behavior inferred from the X-ray spectra of several active
galactic nuclei (AGNs).
%{\em Chandra} X-ray spectra.
%We also compute the detailed profiles of \fexvii~ and \fexxv~
%whose LOS velocities [$v_{\rm los}($\fexvii$) \sim 100 - 300$ km
%s$^{-1}$ at $\log \xi \sim 2-3$ while $v_{\rm los}($\fexxv$)
%\sim 1000 - 4000$ km s$^{-1}$ at $\log \xi \sim 4-5$] are also in
%agreement with the observations.
For the specific wind structure and X-ray spectrum we also compute
detailed absorption line profiles for a number of ions to obtain
their LOS velocities, $v \sim 100 - 300$ km~s$^{-1}$ (at $\log \xi
\sim 2-3$) for \fexvii~ and  $v \sim 1,000 - 4,000$ km~s$^{-1}$ (at
$\log \xi \sim 4-5$) for \fexxv, in good agreement with the
observation. Our models describe the X-ray absorption properties of
these winds with only {\it two parameters}, namely the
mass-accretion rate $\dot m$ and LOS angle $\theta$. The probability
of obscuration of the X-ray ionizing source in these winds decreases
with increasing $\dot m$ and increases steeply with the LOS
inclination angle $\theta$. As such, we concur with previous authors
that these wind configurations, viewed globally, incorporate all the
requisite properties of the parsec scale ``torii" invoked in AGN
unification schemes. We indicate that a combination of the AMD and
absorption line profile observations can uniquely determine these
model parameters and their bearing on AGN population demographics.

\end{abstract}

\keywords{accretion, accretion disks --- galaxies: active ---
methods: numerical --- quasars: absorption lines --- X-rays: galaxies}

\baselineskip=15pt

\section{Introduction}

The issue of accretion as the power source behind Active Galactic
Nuclei (AGNs) was decided as early as 1969 \citep[][]{LyndenB69}.
Since then, a 40-year
effort to unravel the physics underlying this process has produced a
great body of observational and theoretical work covering a host of
the AGN properties and phenomena. However, due to the small angular
size of a black hole horizon, whether at the galactic or extragalactic
setting, these studies were effectively performed mainly in the
spectral (but also the time) domain; the spatial structure of
accretion flows, but those of scale of many parsecs,
was then inevitably delegated to models, whose validity was
determined by their ability to reproduce the spectral and timing
observations with sufficient accuracy. At the risk of
oversimplifying the issue, the spectral properties of AGNs were
basically outlined (among other works) in \citet{Sanders89} who
showed that the Spectral Energy Distribution (SED) of AGNs includes
three broad components in the IR, optical-UV and X-ray bands, with
roughly equal energy per decade, with the radio comprising only a
small fraction of their bolometric luminosity, of order $\sim
10^{-6}$ in the Radio-Quiet (RQ) AGNs and $\sim 10^{-3}$ in Radio-Loud (RL) AGNs.

Considering that accretion produces most of its power in the last 10
or so Schwarzschild radii, $r_S$, outside the black hole horizon,
barring an inherently non-thermal emission process (as it turns out
to be the case with blazars), it is strange that the observed SEDs
exhibit roughly constant luminosity per decade. Interestingly, the
simplest of assumptions, namely radiation of the AGN bolometric
luminosity in black body form by an object (most likely in the form
of a cold disk of $T \sim 10^4 - 10^6$ K) of size a few $r_S$ of a
supermassive black hole ($M \sim 10^6-10^8 \, M_{\odot}$ where
$\Msun$ is the solar mass), implies peak emission at UV frequencies,
consistent with the observationally identified UV feature referred
to above, known as the Big Blue Bump (BBB); as such, this  feature
was modeled with emission by a geometrically-thin, optically-thick
accretion disk extending usually to the Innermost Stable Circular
Orbit (ISCO) of the black hole. Of the other two distinct components
the X-rays are generally attributed to emission by an
optically-thin, hot ($T \sim 10^8-10^9$ K) corona overlying this
disk and heated by the action of magnetic fields that thread the
geometrically-thin disk, in close analogy to the Solar Corona
\citep[e.g.][]{Haardt91,Kawanaka08}. Finally, the IR emission was
accounted for as the result of reprocessing the O-UV radiation by a
geometrically thick (scale height $h$ roughly equal to the local
radius $r$, i.e. $h \sim r$), cool, molecular torus at large
distances ($\gsim 1$ pc) from the central engine [an arrangement
that presents a bit of a problem, since the temperature of this
torus ($\sim 10 - 100$ K) is much less than the virial temperature
of the gas at that distance and would lead to a configuration with
$h \ll r$; see Krolik \& Begelman~(1988).
The region between the torus and the accretion disk was thought to
be occupied by clouds with velocity widths ranging from tens of
thousands to several hundreds of kilometers per second, producing
the observed line radiation, with the ensemble of these components
encapsulated in the well known cartoon of \citet[][]{UP95}.

While in most models these components are generally thought of as
independent, Principal Component Analysis (PCA) of multiwavelength
AGN features by \cite{BorGre92} and \cite{Bor02}  indicated that
most of the variance in the measured optical emission line
properties and a broad range of continuum features (radio, optical,
X-ray) of AGNs was contained in two sets of correlations,
eigenvectors of the correlation matrix. This analysis suggested that
the properties of these features are not independent (despite their
diverse locations and emission processes) but are well correlated by
two single underlying physical parameters through the physics of
accretion and the conversion of its power to radiation and outflows.
On the other hand, general theoretical arguments suggesting, e.g. a
tight correlation between variations in the X-ray and O-UV
components, given their implied proximity, were not confirmed by
observations \citep{Nandra98}.

The advent of X-ray instrumentation and in particular X-ray
spectroscopy, established the presence of spectral components
apparently present across the entire range of compact object masses
from galactic black hole candidates (GBHCs) to luminous AGNs, namely
outflowing X-ray absorbing matter in our line of sight (LOS)
manifested as complex (blueshifted) absorption features of hydrogen
equivalent column density $N_H \sim 10^{21} - 10^{23}$ cm$^{-2}$.
The presence of this absorbing material, often referred to as Warm
Absorber, was first established in {\it ASCA} observations of many
AGNs \citep[e.g.][]{ReyFab95} and GBHCs \citep[e.g.][]{Miller06}
which provided clear evidence of highly ionized oxygen \ovii ~(0.739
keV) and/or \oviii ~(0.871 keV) along the LOS. More recently,
grating spectra of high-resolving power obtained by {\it Chandra}
and {\it XMM-Newton} have enabled the study of these absorption
features in greater detail leading to the conclusion that they are
present in a large fraction of AGNs and span a wide range ($\sim
10^5$) in ionization parameter \citep[e.g.][hereafter HBK07]{HBK07}.
Some sources appear to contain low and/or high velocity outflows
\citep[e.g. HBK07;][]{HBA09} while a handful of objects exhibit a
trans-relativistic outflow
\citep[e.g.][]{Chartas09,Pounds09,Reeves09}. In addition in both AGNs
and GBHCs the observed outflows from at least a number of sources
have been suggested to be accelerated magnetically (rather than by
another process) off an accretion disk (e.g., see Miller et
al.~2006, 2008 for GRO~J1655-40; Kraemer et al.~2005 and Crenshaw \&
Kraemer~2007 for NGC~4151). A brief review of luminous AGN winds is
found in \citet{Brandt09}. These facts imply the presence of gas
covering a large fraction of the solid angle and distributed over a
large range of radii, perhaps the entire range between the X-ray
source and the torus, indicating AGNs to be multiscale and
multiwavelength objects rather than a class with properties
determined only by the power released by accretion in the black hole
vicinity. This last point has gained further support with the
discovery of the correlation between the accreting black hole mass
and the velocity dispersion of the surrounding stellar population
\citep{FerMer}.

From the theoretical point of view, most early AGN treatments were
focused on the structure of the innermost regions of the accretion
flow and as such they limited themselves to the study of the
conditions in a rather narrow range of radii. Nonetheless, several
treatments of magnetohydrodynamic (MHD) accretion disk winds, aiming
to account for the observed AGN outflows, employed self-similar
solutions, which naturally span a large range in radius (e.g.
Blandford \& Payne 1982, hereafter BP82; Contopoulos \& Lovelace
1994, hereafter CL94). Using the structure of these solutions,
\citet[][hereafter KK94]{KK94} proposed that the so-called molecular
torus of the AGN unification scheme is a dynamical rather than a
static object, namely an MHD wind of the type suggested in the above
works. KK94 also contented that agreement between model and
observation demanded a rather specific type of wind, a particular
case of those discussed in CL94, one that we also concentrate on in
the present work. At the same time, they noted that the 2D geometry
of these winds (with low column at inclination angles $\theta \simeq
0^{\circ}$, i.e. along the wind axis, and high column at $\theta
\simeq 90^{\circ}$) provided, in addition, a natural framework for
the unification scheme of type I and II Seyfert galaxies proposed by
Antonucci \& Miller~(1985), thereby linking the large scale AGN
spectral classification to the dynamics of AGN accretion/outflows.

Self-similar solutions of the accretion flow equations were also
provided by \citet{NY94,NY95a}, who considered the structure of hot
accretion flows in the regime of low accretion rate, i.e. for
accretion rates less than the Eddington value, $\dot m \equiv \dot
M/ \dot M_{\rm E} \lsim 1$. Because for these low rates the
accretion time scale $\tau_{acc}$ is shorter than the gas cooling
time $\tau_{\rm cool}$, in fact $\tau_{\rm acc} \simeq \dot m \,
\tau_{\rm cool}$, only a fraction $\dot m$ of the energy released in
the dissipation of the plasma kinetic energy is radiated away with
the remainder advected into the black hole (hence the term Advection
Dominated Accretion Flows or ADAFs); as a result, the accretion
luminosity is then proportional to $\dot m^2$ (rather than $\dot
m$). The high temperature of the inner ADAFs, provides naturally for
the hot component required to produce the observed X-ray emission,
which, rather than being an independent component, within ADAFs is
incorporated in the dynamics of accretion itself \citep[see][for
recent review]{Narayan08}. More recently, \citet{BlandBegel},
elaborated further on ADAFs elucidating, among others, the reason
for which the Bernoulli integral of ADAFs is positive \citep[a point
that had been noted by][]{NY94,NY95a}, that being that the viscous
stresses transfer outward, in addition to angular momentum, also
mechanical energy. They then proposed that this excess energy can be
carried away in the form of wind over a large range of radii,
thereby leading to configurations similar to those obtained in the
more detailed studies of BP82 and CL94, while at the same time
maintaining some of the general properties of ADAFs. The combination
of radiatively inefficient flows with the simultaneous presence of
winds led to the nomenclature Advection Dominated Inflow-Outflow
Solutions (ADIOS). Following these pioneering works, many attempts
have been made in recent years to reproduce and explain the
kinematics and X-ray spectra of the observed outflows in AGNs
\citep[see, among
others,][]{Proga03,Everett05,Sim05,SD07,DKP08,Sim08,SD09}.
%which are not
%different in their wind properties from the detailed models of
%CL94.

Motivated by these considerations we are taking a closer look at the
issue of Warm Absorbers (X-ray absorbing medium) in terms of
specific disk-wind models, namely those of CL94. As it will be
discussed in more detail later on, these solutions are characterized
by a parameter $q$, which determines the distribution of axial
current in the wind as a function of the radius; this parameter, in
conjunction with the MHD conservation laws and the radial force
balance, determines both the radial dependence of the matter density
in the wind and the radial dependence of the toroidal magnetic field
component $B_{\phi}$. Taking our lead from the works of
\citet{Behar03} and HBK07, in the present work we restrict our
attention to wind models with $q=1$, i.e. the value used also by
KK94. This is in a sense a critical value of this parameter because,
as discussed in CL94, it leads to winds with toroidal magnetic field
$B_{\phi} \propto 1/r$
%($R$ is the cylindrical radial coordinate)
and therefore magnetic field energy on the disk that diverges only
logarithmically at small and large $r$ and as such it can be
considered as a ``minimum magnetic energy" configuration (KK94);
values of $q \neq 1$ lead to configurations with power-law
divergence either at small or large $R$. In addition, the choice
$q=1$ leads to a wind with radial density profile $n(r) \propto
r^{-1}$, i.e. a profile with equal column per logarithmic radial
interval (and normalization that depends on the polar angle $\theta$
and the mass-accretion rate $\dot m$), a fact that is in general
agreement with the profiles implied by the ionized absorbers data
analysis to date (HBK07). It is worth noting that a similar density
profile was invoked by \citet{KHT96} and \citet{HKT97} to explain
the Fourier frequency dependence of the soft-hard X-ray lags
observed in GBHCs and in AGNs (Papadakis et al. 2002).

With the wind 2D ($r$ and $\theta$) density field as a function of
radius and angle provided by the models of CL94 (i.e. ignoring at
present effects that can be attributed to other agents that could
affect the wind structure) and the assumption of a point X-ray
source of a given spectrum at the origin, one can compute their 2D
ionization structure and in particular the local column density of
specific ionic species as a function of radius and angle. One can
also compute the integrated column of each such ion, quantities that
can be directly compared to observation (HBK07). This we do in the
present work. { While a number of issues remain open or are
sidestepped in the present self-similar wind models, this is the
price to pay in order to limit the number of free parameters to just
two (the wind mass flux and the observer inclination angle), a fact
that allows the study of the properties of warm absorbers within the
global perspective of AGN unification}.

Our paper is structured as follows: in \S 2 we review the MHD wind
equations and structure as given in CL94 and make a connection of
these winds with flows of the ADAF/ADIOS type by relating the wind
density normalization to the ionizing luminosity produced by the
idealized point source at the coordinate origin. In \S 3 we provide
the general 2D ionization structure, the local column of
specific ions as a function of radius as well as the total ionic
column of a given charge state and for each element, as well as the
absorption line profiles of selected transitions. In \S 4 the results
are compared to observation, the properties and limitation of the
present treatment are discussed and a course for future work is charted.

%\subsection{X-ray Photoionization \& Atomic Physics}

\section{Description of the Model}

Our wind model consists primarily of two parts: ($i$) The
outflow/wind structure originating from a geometrically-thin
accretion disk as described by CL94 and ($ii$) The computation of
the ionization of the wind plasma under local heating-cooling and
ionization equilibrium, by employing the photoinization code
\verb"XSTAR" \citep{KB01}. These two steps are self-consistently
coupled in a scheme similar to some previous work in a slightly
different context \citep[e.g.][]{SD07,DKP08,SD09}. All calculations
are performed under steady-state, axisymmetric conditions as shall
be described in detail in the following sections.

\subsection{Self-similar MHD Wind from Accretion Disks}

While the detailed formalism of self-similar MHD wind solutions is
found in CL94, let us here briefly summarize some
of their characteristics. The basic equations of this problem are
those of steady-state, nonrelativistic, ideal MHD that include
gravity and gas pressure, namely
\begin{eqnarray}
\nabla \cdot (\rho \bf{v}) = 0 & &  \textrm{(mass conservation)} \ ,
\label{eq:MHD-1}
\\
\nabla \times \bf{B} = \frac{4 \pi}{c} \bf{J} & &
\textrm{(Ampere's law)} \ , \label{eq:MHD-2}
\\
\bf{E}+ \frac{\bf{v}}{c} \times \bf{B} = 0 & &  \textrm{(ideal
MHD)} \ , \label{eq:MHD-3}
\\
\nabla \times \bf{E} = 0 & &  \textrm{(Faraday's law)} \ ,
\label{eq:MHD-4}
\\
\rho (\bf{v} \cdot \nabla) \bf{v} = -\nabla p - \rho \nabla\Phi_g
+ \frac{1}{c} (\bf{J} \times \bf{B}) & &  \textrm{(momentum
conservation)} \ , \label{eq:MHD-5}
\\
\nabla \cdot \bf{B} = 0 & &  \ . \label{eq:MHD-6}
\end{eqnarray}
Here, the plasma wind velocity $\bf{v}$ is assumed to be frozen
into the global magnetic field $\bf{B}$; $\bf{E}$ is the electric
field; $\bf{J}$ is the electric current density; and $c$ is the
speed of light. The gas pressure $p$ and mass density $\rho$ of
the wind are related through an equation of state
\begin{eqnarray}
p &=& K \rho^\Gamma \ , \label{eq:eos}
\end{eqnarray}
where $K$ and $\Gamma$ are the adiabatic constant and adiabatic
index respectively.
%We set $\Gamma=5/3$ throughout this paper.
The assumption of axial symmetry allows one to separate poloidal
quantities from toroidal ones as $\bf{B} = \bf{B}_p + \bf{B}_\phi$
and $\bf{v} = \bf{v}_p + \bf{v}_\phi$, with
\begin{eqnarray}
\bf{B}_p &=& \frac{1}{r \sin\theta} \nabla \Psi \times \hat{\bf{\phi}}
\ , \label{eq:eos}
\end{eqnarray}
where, $\Psi(r,\theta)$ is the magnetic flux function. We will
henceforth work in spherical coordinates ($r, \theta, \phi$) [note
that CL94 used cylindrical ones ($R,\phi,Z$)]. Magnetic field
lines lie along surfaces of constant value of $\Psi$. Furthermore,
under conditions of steady-state, the following quantities are
conserved along field lines characterized by the value of $\Psi$
\begin{equation}
4\pi \rho \frac{v_p}{B_p} \equiv F(\Psi) \ , \label{eq:F}
\end{equation}
\begin{equation}
\frac{1}{r \sin\theta} \left(v_\phi-\frac{F}{4\pi \rho} B_\phi \right) \equiv
\Omega(\Psi) \ , \label{eq:Omega}
\end{equation}
\begin{equation}
r \sin\theta \left(B_\phi - v_\phi F \right) \equiv H(\Psi) \ , \label{eq:H}
\end{equation}
\begin{equation}
\int_{\Psi=const} \left(d\rho/\rho \right) + \frac{1}{2} v^2 +
\Phi_g - r \sin\theta v_\phi \Omega \equiv J(\Psi) \ , \label{eq:J}
\end{equation}
\begin{equation}
k_B \left(\Gamma -1 \right)^{-1}\ln(K) \equiv S(\Psi) \ , \label{eq:S}
\end{equation}
where $F(\Psi)$ is the ratio of mass to magnetic fluxes,
$\Omega(\Psi)$ is the angular velocity of magnetic field lines,
$H(\Psi)$ corresponds to the specific angular momentum (it
includes magnetic torques), $J(\Psi)$ is the specific energy
(Bernoulli) integral, and $S(\Psi)$ is the specific entropy. Also, $k_B$
is Boltzmann's constant, and $\Phi_g$ is the Newtonian
gravitational potential. The wind velocity field $\bf{v}$ is
related to the magnetic field $\bf{B}$ as
\begin{equation}
{\bf v} = r \sin\theta \Omega \hat{\phi} + \frac{F}{4\pi\rho} {\bf
B} \ . \label{eq:eos}
\end{equation}
CL94 first showed that one can obtain general self-similar
solutions of the above system of equations of the form
\begin{eqnarray}
\bf{B}(r,\theta) &\equiv& (r/r_o)^{q-2} {\bf b}(\theta)B_o \ ,
\label{eq:eos} \\
\bf{v}(r,\theta) &\equiv& (r/r_o)^{-1/2} {\bf v}(\theta)v_o \ ,
 \label{eq:eos} \\
     p(r,\theta) &\equiv& (r/r_o)^{2q-4} {\cal P}(\theta)B_o^2 \ ,
 \label{eq:pres2}  \\
\rho(r,\theta) &\equiv& (r/r_o)^{2q-3} {\cal R}(\theta)B_o^2v_o^{-2}
\ .
\end{eqnarray}
Here, $r_o$ is the launch radius of a characteristic field/flow
line at the base of the wind at $\theta=90\degr$ while $B_o$ and
$v_o$ are respectively the magnitude of the vertical components of
the magnetic field and the initial rotational wind velocity at the
same foot point; $\bf{b}(\theta)$ and $\bf{v}(\theta)$ are
dimensionless with $b_z(90\degr)=v_\phi(90\degr) \equiv 1$ while
${\cal P}(\theta)$ and ${\cal R}(\theta)$ denote the angular
dependence of the pressure and wind density, respectively. The
exponent $q$ is a free parameter that controls the radial scaling
of the magnetic field. One sees directly that $q$ discriminates
current-carrying wind configurations (when $q> 1$, $rB_\phi$ grows
with $r$; the wind carries a net axial current) from
zero-total-current ones (when $q<1$, $rB_\phi\rightarrow 0$ for
$r\rightarrow \infty$; the wind carries a singular axial current
and its corresponding return current; see CL94 and references).
Obviously,
\begin{eqnarray}
\Psi(r,\theta) &=& (r/r_o)^q \psi(\theta)\Psi_o
  \ , \label{eq:eos}
\end{eqnarray}
where $\psi(\theta)$ is dimensionless, $\psi(90^{\circ})=1$, and
$\Psi_o \equiv B_o r_o^2 b_z(90^{\circ})q^{-1}$ is the poloidal magnetic
flux through radius $r_o$. The above scalings allow us to obtain the
geometry of the poloidal field/flow lines (characterized by the
value of $\Psi$) as
\begin{eqnarray}
\left.r\right|_\Psi (\theta) & \equiv & r_o (\Psi
/\Psi_o)^{1/q}\psi(\theta)^{-1/q} . \label{eq:geom}
\end{eqnarray}
Note that the function $\psi(\theta)$ characterizes the shape of
all field/flow lines. That is, all field/flow lines are
self-similar to the `characteristic' field/flow line that
originates at $r=r_o$ and $\theta=90^o$ on the disk.
Self-similarity further implies that
\begin{eqnarray}
F(\Psi) &=& (\Psi/\Psi_o)^{1-3/2q}F_o B_o v_o^{-1} \ , \\
\Omega(\Psi) &=&  (\Psi/\Psi_o)^{-3/2q}\Omega_o v_o r_o^{-1} \ , \\
H(\Psi) &=&  (\Psi/\Psi_o)^{1-1/q}H_o B_o r_o \ , \\
J(\Psi) &=&  (\Psi/\Psi_o)^{-1/q}J_o v_o^2 \ ,
\end{eqnarray}
\begin{equation}
S(\Psi) = k_B (\Gamma-1)^{-1} [\{(2-4/q)-\Gamma (2-3/q)\} \ln
(\Psi/\Psi_o) + \ln \cal{K}] \ , \label{eq:entropy}
\end{equation}
where ${\cal K}={\cal P}(\theta)/{\cal R}(\theta)^\Gamma$ is a
dimensionless adiabatic constant. The adiabatic speed of sound
scales as
\begin{eqnarray}
c_s(r,\theta) \equiv \left(\frac{\partial p}{\partial
\rho}\right)^{1/2}_{\Psi=const.} = (\Psi/\Psi_o)^{-1/2q}
\left[{\cal K}\Gamma \, {\cal R(\theta)}^{\Gamma-1}\right]^{1/2}
v_o = (\Psi/\Psi_o)^{-1/2q} c_s(\theta)v_o \ . \label{eq:sound}
\end{eqnarray}
Finally, one can combine the dimensionless forms of
equations~(\ref{eq:F})-(\ref{eq:entropy}) that involve the
dimensionless quantities $( {\bf v}, {\bf b}, {\cal P}, {\cal R},
c_s )$ as functions of $\theta$, and thus express the poloidal
projection of the momentum balance equation~(\ref{eq:MHD-5}) [the so
called generalized Grad-Shafranov (GS) equation] in the form of a
second-order ordinary differential equation for $\psi(\theta)$ as
\begin{eqnarray}
\psi''(\theta) = f(\theta,\psi,\psi') \ . \label{eq:GS}
\end{eqnarray}
This equation (too complicated to write down explicitly in this
paper) is integrated numerically from the surface of the disk (taken
for simplicity to lie at $\theta=90\degr$) to axial infinity
($\theta=0\degr$). The initial conditions are $\psi(90\degr)=1$ and
$\psi'(90\degr)$, essentially a free-parameter that depends
on the details of the internal disk structure, which, for a given
value of $H_o$, is adjusted in order for the wind to cross the
\Alfven point (the point at which the the wind poloidal speed $v_p$
becomes equal to the local \Alfven speed - see CL94 for a detailed
description) continuously and smoothly. According to
equation~(\ref{eq:geom}), the form of the solution $\psi(\theta)$
characterizes the poloidal shape of the field lines, and in
particular, the initial condition $\psi '(90\degr)$ characterizes
the angle the poloidal field/flow lines make with the disk at their
foot points [readers can easily verify that this angle is equal to
$\tan^{-1}\{q/\psi'(90\degr)\}$].

To numerically solve the above wind equation~(\ref{eq:GS}) we assume
that the plasma is initially in Keplerian rotation $v_o \equiv
v_{\phi}(90\degr) = v_{K}$ at the disk surface $(\theta = 90\degr)$
with only a small out-of-the-disk velocity component ($v_r, v_\theta
\ll v_{\phi}$ at $90\degr$). As a fiducial solution we fix the
following set of parameters as described in CL94; $q=1$, $F_o =
0.15, H_o = -1.584$ with kinematic variables $v_\phi(90\degr)=1$,
$v_\theta(90\degr)=0.01$, ${\cal K} = 0.01$ and $\Gamma=5/3$. With
these initial conditions, the remaining conserved quantities are
calculated as $\Omega_o = 1-v_\theta(90\degr)(F_o+H_o)=0.983$, and
$J_o = v_\theta^2(90\degr)[1+(\psi '(90\degr))^2]/2-1/2-\Omega_o\sim
-1.483$ (see CL94 for details). The only freedom left is the value
of $\psi '(90\degr)$ that is obtained iteratively so that the wind
passes smoothly through the \Alfven surface. This important physical
constraint yields $\psi '(90\degr) = 0.98091$. Note that for
sufficiently small $F_o$ (a magnetically dominated flow) its
velocity is nearly Keplerian, setting the value of $\Omega_o$, while
passing through the Alfv\'en point fixes $H_o$ and effectively the
remaining constants of the flow, with $q$ as the one meaningful free
parameter.
%
%-----------------------------------------------Place Figure~1
\begin{figure}[t]% ------------------------------------- Figure~1
\epsscale{0.99} \plottwo{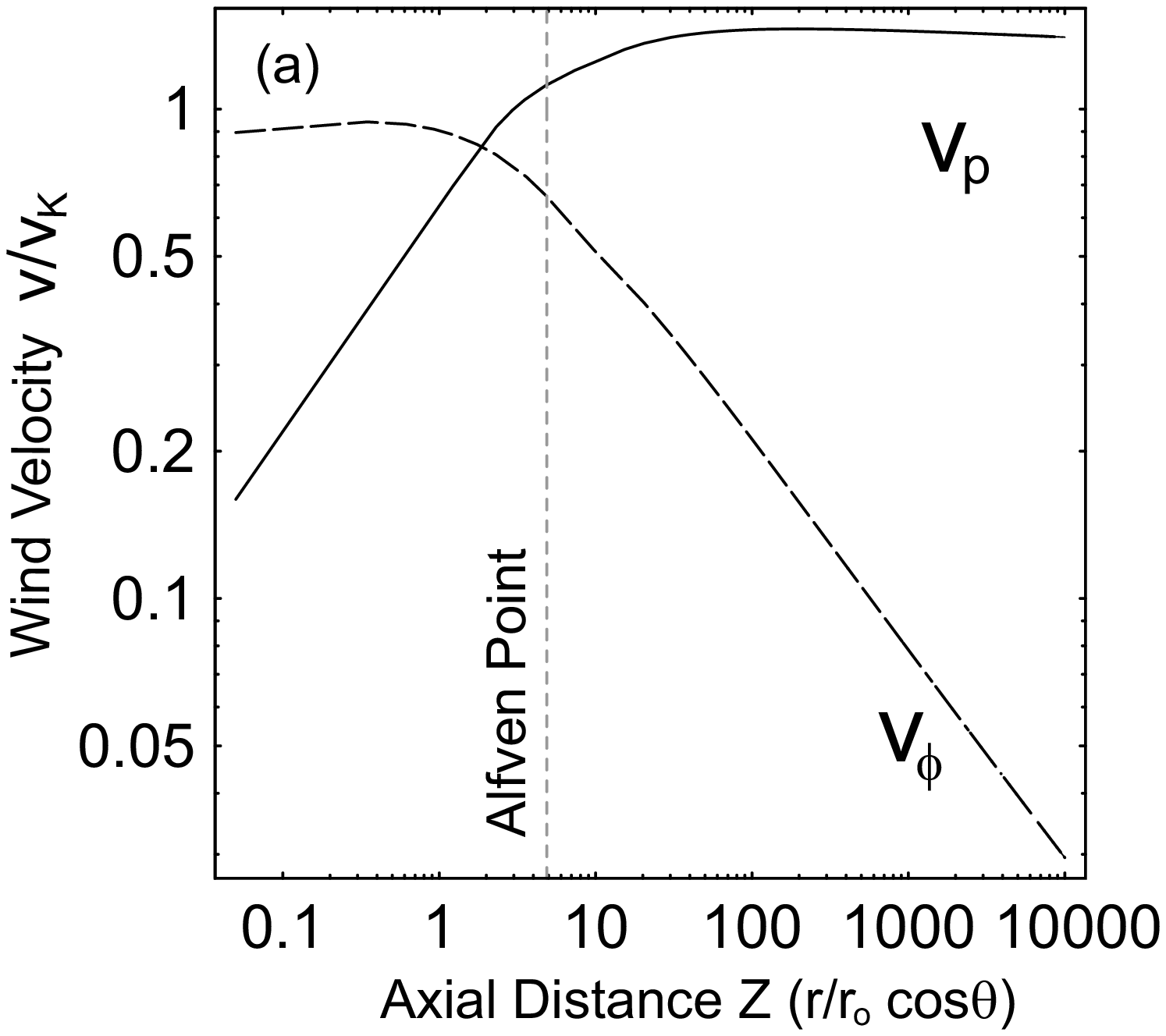}{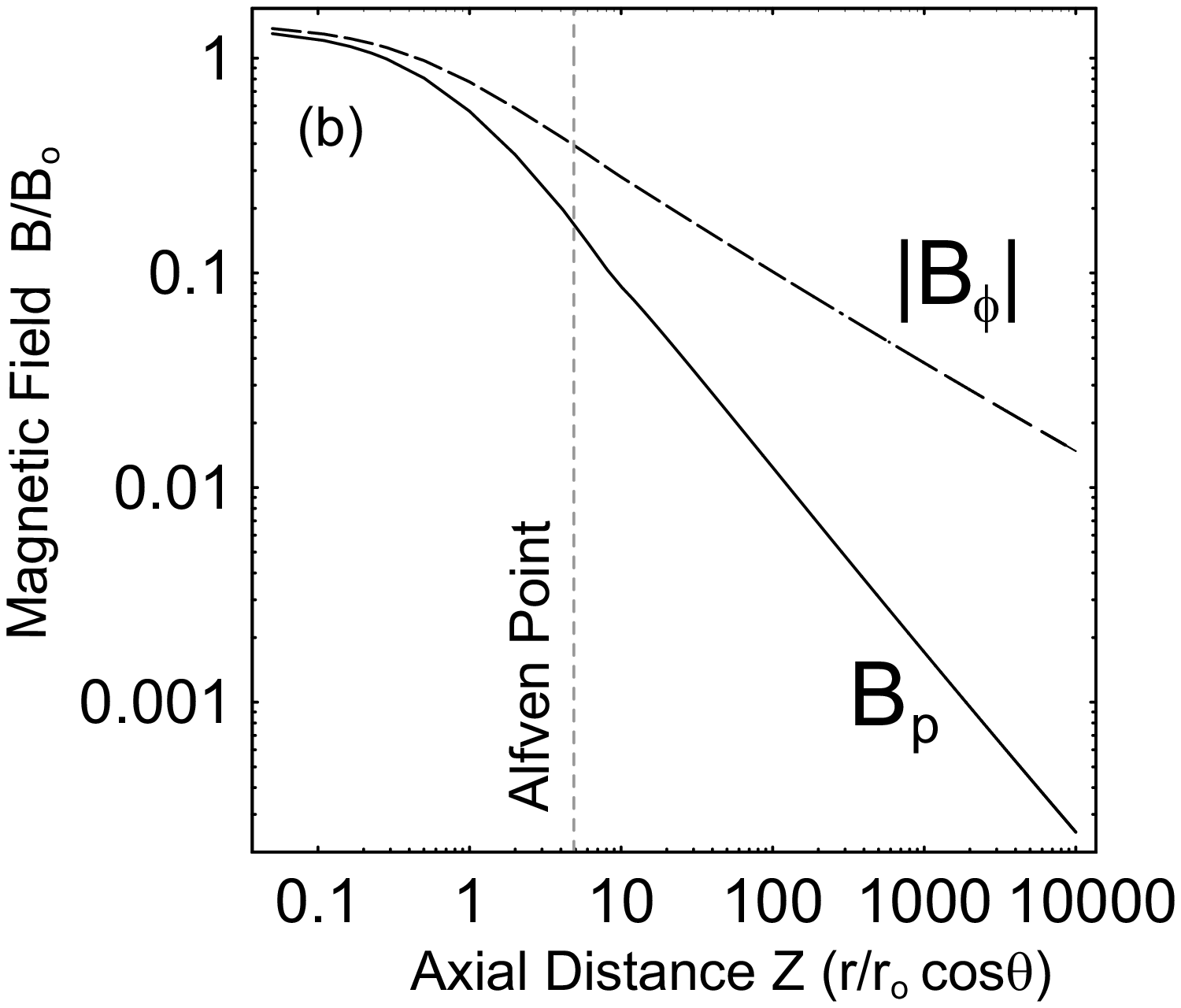} \caption{MHD wind
solution of the GS-equation~(\ref{eq:GS}) in the poloidal plane: (a)
poloidal $v_p$ (solid) and toroidal $v_\phi$ (dashed) velocities (in
units of $v_o = v_K$) as a function of axial distance $Z \equiv
(r/r_o) \cos \theta$ and (b) poloidal $B_p$ (solid) and toroidal
$|B_\phi|$ (dashed) magnetic fields (in units of $B_o$) also as a
function of $Z$ along the characteristic field/flow line originating
at $r=r_o$ on the disk for the parameters described in the text.
Note that we show the absolute value of $B_\phi(<0)$. Vertical
dotted lines denote the axial location of the \Alfven point. }
\label{fig:wind}
\end{figure}
%-------------------------------------------------------------
%
The obtained wind kinematics are shown in Figure~\ref{fig:wind}a
and the magnetic field components in Figure~\ref{fig:wind}b. As
discussed in CL94, different values of the exponent $q$ allow for
distinct streamline geometries in the poloidal plane, which in
turn is associated with other observable quantities of the wind
(such as the density profile; see following sections). As is
initially set up, this wind is launched from a thin Keplerian disk
with only a small out-of-the-disk surface velocity component.
The magnetic field there is dragged by the rotating plasma
($v_\phi > 0$) resulting thus in a negative toroidal field
component ($B_\phi <0$).
As the wind leaves the disk, magnetic torques act on the outflowing
plasma and magnetic energy is efficiently converted into kinetic
energy of the wind along the symmetry axis. The magnetic field also
plays an important role in collimating the wind at large distances.
In this particular case the wind is found to end up with terminal
velocity roughly a few times the initial rotational speed and remains below
fast magnetosonic speed.

The astute reader may note that X-ray heating is likely to produce
thermal velocities inconsistent with the self-similarity of the
sound speed invoked above (i.e. $c_s^2 \propto 1/x$; see Fig.~\ref{fig:amd}). We will discuss this in \S 3.1.

Knowing the wind streamline described by equation~(\ref{eq:GS}) one
can separately write down radial and angular dependence of the wind
number density $n(r,\theta)$ as %%
%\begin{eqnarray}
%n(r,\theta;\eta_w,\dot{m}) \equiv n_0  \eta_w \frac{\dot{m}}{\hat{M}}
%\left(\frac{r_S}{r}\right)^{3-2q} {\cal N}(theta) = {\cal N}(theta) \eta_w \; n_0
%x^{-3+2q}\frac{\dot{m}}{M}\ , \label{eq:n}
%\end{eqnarray}
%
%
\begin{equation}
n(r,\theta) \equiv \frac{\rho(r,\theta)}{\mu m_p}=n_o x^{2q-3}{\cal
N}(\theta)
%n_0  \eta_W \left(\frac{r_S}{r}\right)^{3-2q} {\cal N}(\theta) =
%\eta_W{\cal N}(\theta) \, \frac{\dot m}{2 \sigma_{\rm T} r_S}
%x^{2q-3}
\ , \label{eq:n}
\end{equation}
where $n_o \equiv B_o^2 {\cal R}(90\degr)/(\mu m_p v_o^2)$ is the
number density normalization at the initial characteristic launching radius
$r_o$, ${\cal N}(\theta) \equiv {\cal R}(\theta)/{\cal
R}(90\degr)$ is the normalized angular dependence of the wind
number density, and $x \equiv r/r_o$ is the non-dimensional radial
coordinate; $m_p$ is the proton mass and $\mu=1.26$ the mean molecular
weight of the wind. In what follows, we will take the
characteristic radius to be equal to the Schwarzschild radius,
namely $r_o=r_S$, taken to be the inner edge of the accretion flow
with $r_S \simeq 3 \times 10^5 \, \hat{M}$ cm where $\hat{M}
\equiv M/\Msun$ is the mass of the black hole in units of solar
mass $\Msun$. It is well known that free-fall spherical accretion
at the Eddington rate ($\dot m =1$) produces Thomson opacity one
($\tau_{\rm T} \simeq 1$) at the Schwarzschild radius ($r \simeq
r_S$). Using this scaling we can express our wind density
normalization at the inner edge of the disk as
\begin{equation}
n_o = \frac{\eta_W {\dot m}}{2\sigma_T r_S} \ , \label{eq:dnorm}
\end{equation}
where, $\eta_W$ is the ratio of of the mass-outflow rate in the wind
to the mass-accretion rate $\dot{m}$, assumed here to be
%fraction of the total mass flow that goes
%to the wind, typically,
$\eta_W \simeq 1$, and $\sigma_T$ is the Thomson cross-section. It
is important to note here that because the mass flux in these winds
depends in general on the radius, a normalized parameter used
throughout this work, $\dot m$, always refers to the mass flux at
the innermost value of the flow radius, i.e. at $x \simeq 1$.

\begin{figure}[t]% ------------------------------------- Figure~2
\epsscale{0.99} \plottwo{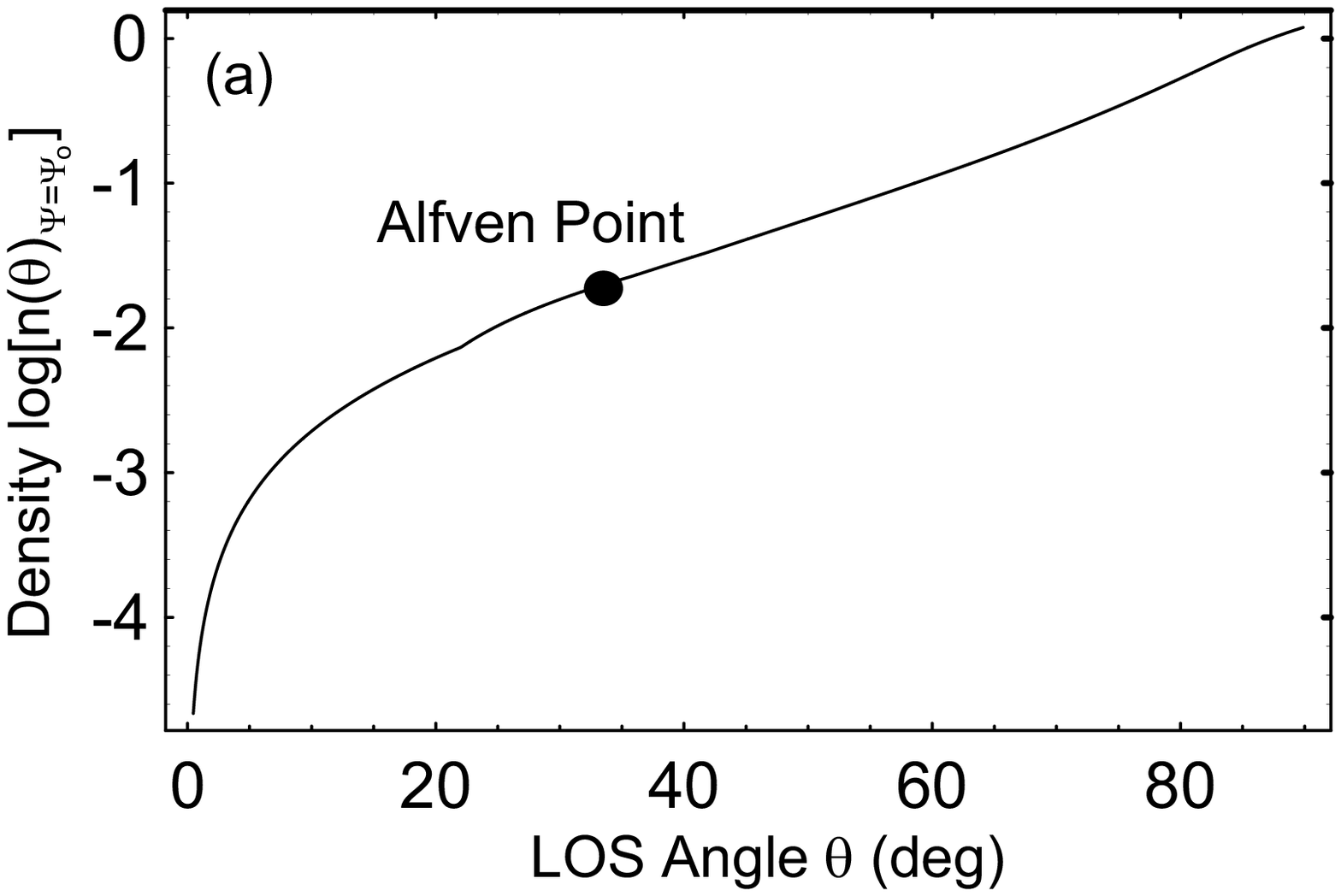}{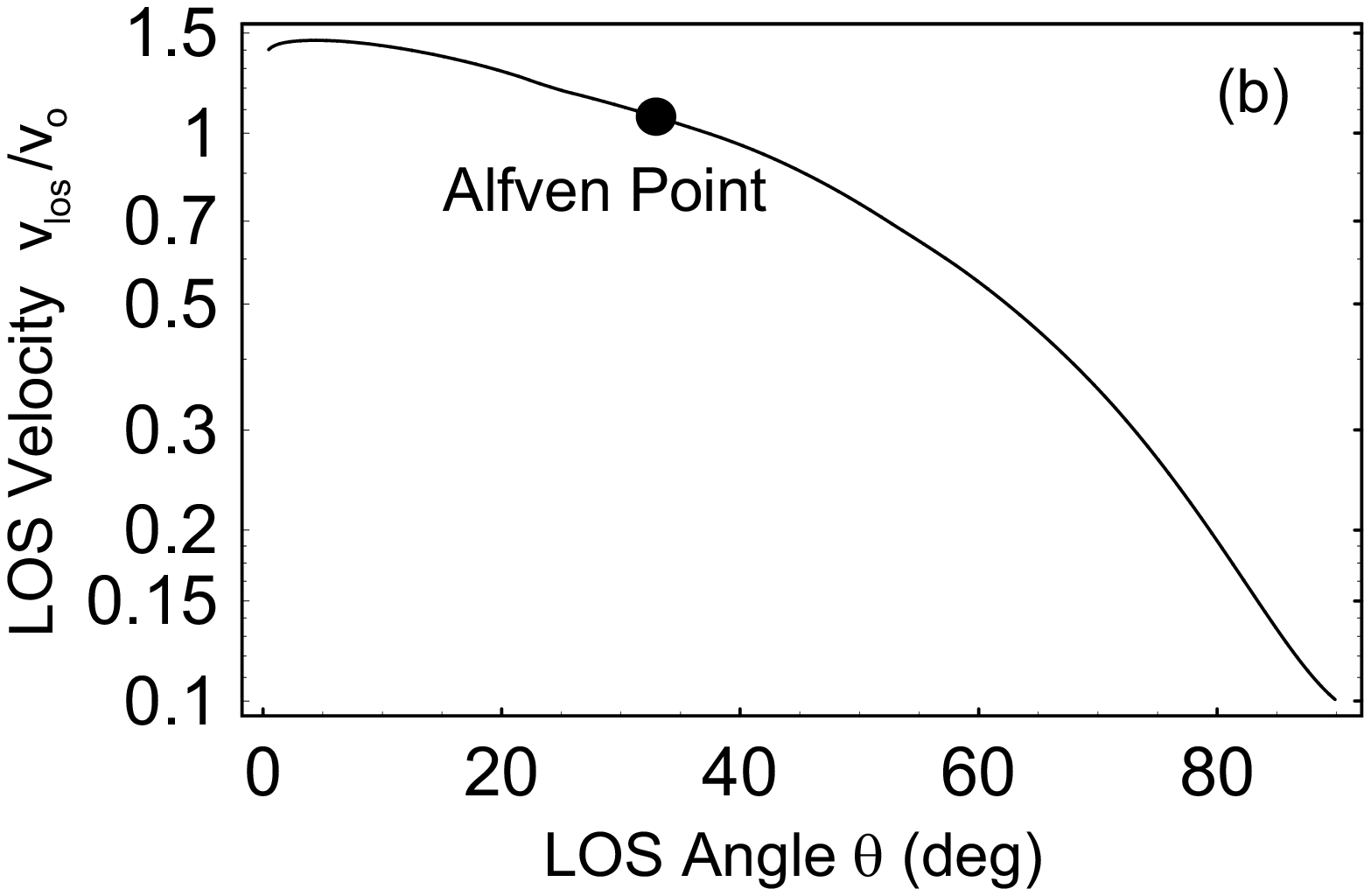} \caption{Wind
properties along a streamline (of $\Psi=\Psi_o$) as a function of
LOS angle $\theta$: (a) density profile $n(\theta)_{\Psi=\Psi_o}$
(normalized to unity) given by equation~(\ref{eq:GS}). (b) LOS wind
velocity $v_{\rm los}$ along the same characteristic field/flow line
(in units of $v_o = v_K$ at the base of the wind). A dot denotes the
position of the \Alfven point. The parameters are the same as in
Figure~\ref{fig:wind}. } \label{fig:density}
\end{figure}

Figure~\ref{fig:density}a exhibits a LOS angle ($\theta$) dependence
of the wind density profile $n(\theta)_{\Psi=\Psi_o}$ from the pole
($\theta = 0\degr$) to the equator ($\theta = 90\degr$) along a
characteristic stream line of $\Psi=\Psi_o$, normalized to unity at
its highest value at $\theta=90^{\circ}$, as obtained from the
self-similar solution of Figure~\ref{fig:wind}. This distribution
illustrates the significant change of the LOS density (and also
column density) with inclination angle (over three decades between
$5\degr \lesssim \theta \lesssim 90\degr$). The LOS wind velocity
$v_{\rm los}$ (in units of $v_o = v_K$) along a characteristic
streamline is shown in Figure~\ref{fig:density}b. We see that
$v_{\rm los}$ increases as the LOS angle $\theta$ becomes smaller
(toward the polar region) because the wind is magnetocentrifugally
accelerated along the way, and furthermore, the direction of $v_{\rm
los}$ becomes more and more parallel to the total wind velocity
${\bf v}$ as well [e.g. $v_{\rm los}(30\degr) \sim 2 v_{\rm
los}(60\degr)$].

%-------------------------------------------------------------

In terms of the scalings of our wind model, the total equivalent
hydrogen column density $N_H$ of wind over the LOS length scale of
$\Delta r$ is given by
\begin{eqnarray}
N_H(\Delta r,\theta) &\equiv& \int_{\Delta r}
n(r,\theta) dr \nonumber \\
&=&  \eta_W {\cal N}(\theta)\, \frac{\dot{m}}{2 \sigma_{\rm T}}
\times
 \left\{
\begin{array}{llr}\frac{1}{2(q-1)}
  \left.x^{2(q-1)}\right|_{\Delta x} &
~~~{\rm if}~~~ q \ne 1 ,  \\
 \left. \ln x\right|_{\Delta x} & ~~~{\rm if}~~~ q=1 , \\
\end{array} \right. \label{eq:column1}
\end{eqnarray}
where the wind is considered to extend from an inner radius
$r=r_{\rm in}=r_o \simeq r_S$ over a length scale of $\Delta r
\equiv \Delta x \cdot r_S$ along the LOS. One should note that the
$q=1$ configurations have the interesting property of equal column
density per decade of radius, i.e. $d N_H / d \log r= \rm{const}
\propto {\cal N}(\theta)$ [as can be seen from
Fig.~\ref{fig:density}a, $n(\theta)_{\Psi=\Psi_o} \simeq
e^{-(\theta-90\degr)/13\degr}$ is an excellent fit to the angular
dependence of the density for $\theta \gsim 10 \degr$ where $\theta$
is measured in degrees]. For $\dot m \simeq 1$ and $\theta \simeq
90\degr$ these winds have a column density of $\simeq 10^{24}$
cm$^{-2}$ per decade of radius. As a result, configurations that
extend over a number of decades in radius can be quite
optically-thick in these directions even for $\dot m \ll 1$.

%\subsection{X-ray Photoionization \& Atomic Physics}

%\subsection{X-ray Photoionization \& Atomic Physics}

\subsection{The Wind Ionization Structure }

With the poloidal field configuration obtained by solving
equation~(\ref{eq:GS}) and the corresponding wind kinematics, we now
are in a position to consider its ionization structure assuming the
presence of a point-like X-ray source at the origin. To this end we
employ the photoionization code \verb"XSTAR" version 2.1kn9, which
solves simultaneously for the ionization balance and thermal
equilibrium of the locally illuminated wind.

The implementation of the above code involves a number of issues and
leads to certain wind observables, amongst which are the following:
($i$) The ionization is locally determined by the photon flux per
electron flux; a proxy for that is the ionization parameter $\xi
\equiv L/(n r^2)$, with $L$ being the luminosity of the ionizing
source; ($ii$) The computation of the ionizing flux requires the
transfer of radiation from the source through the LOS wind, given
that part of it can be removed by scattering and absorption; ($iii$)
Given that our wind models produce, in addition to the density
$n(r,\theta)$, also the wind velocity field $\bf{v}(r,\theta)$ we
are then able to compute also detailed absorption line profiles;
($iv$) The broad range of the wind density with radius leads to ions
with very diverse ionization states, whose distribution per
logarithmic $\xi$ interval, the AMD, discussed in HBK07 and
\citet[][]{Behar09}, can be used to determine the wind radial
profile. We will discuss each one of these issues in detail below.

\subsubsection{The Ionization Parameter}
\begin{figure}[t]% ------------------------------------- Figure~3
\epsscale{0.99} \plottwo{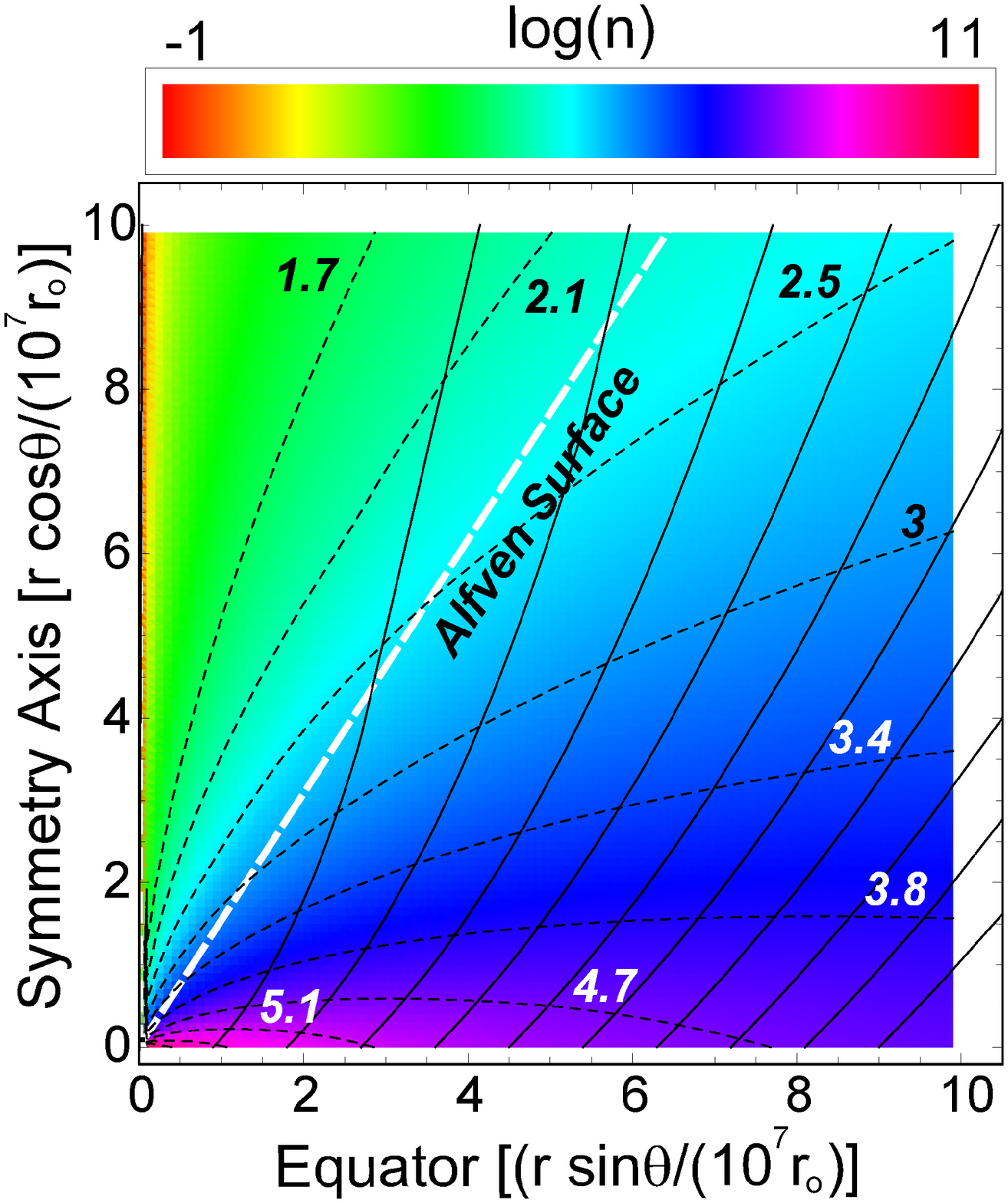}{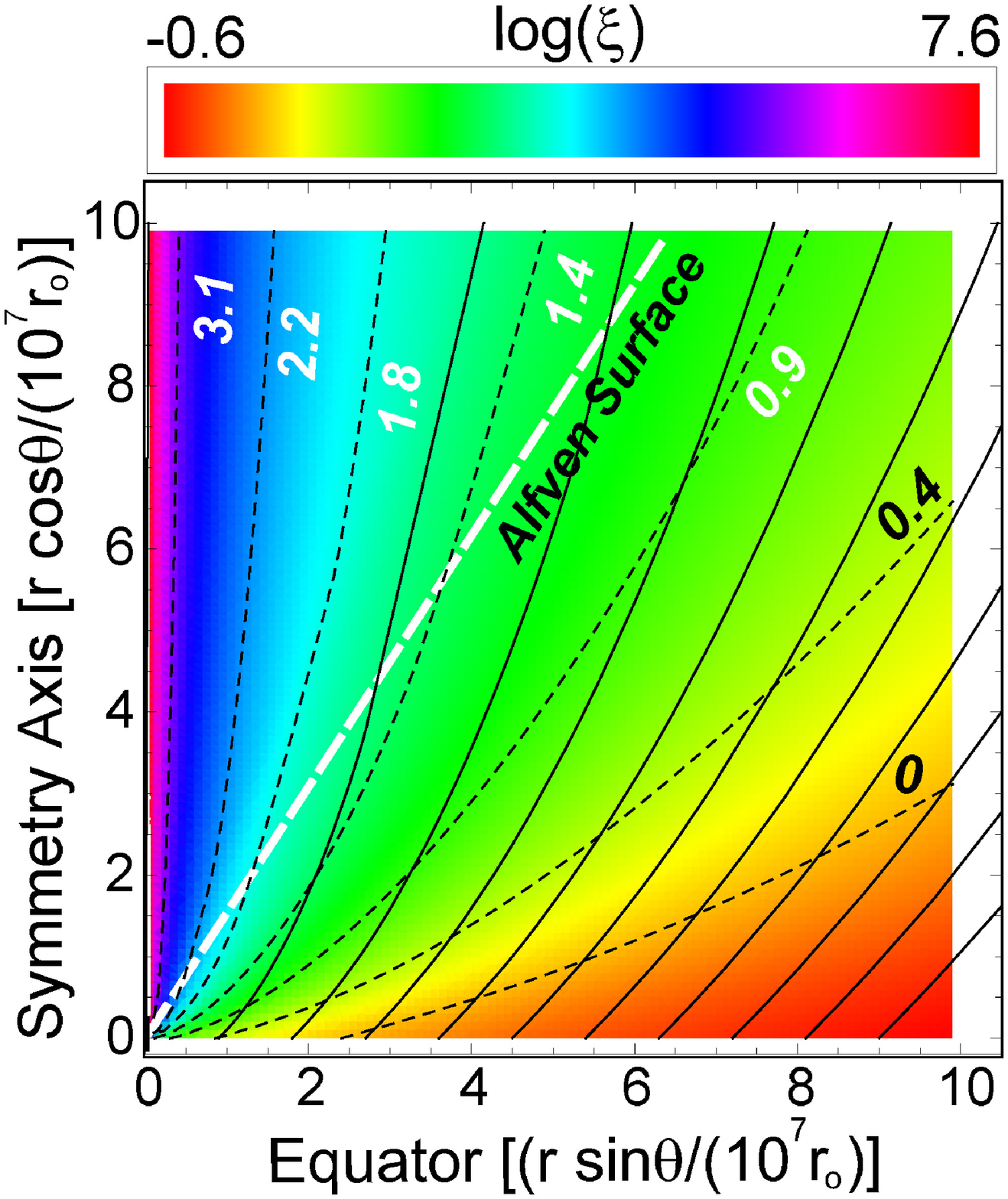} \caption{Poloidal
distribution of (a) the wind density (in units of cm$^{-3}$) $\log n(r,\theta)$ and (b) the
ionization parameter (in units of erg~cm~s$^{-1}$) $\log \xi(r,\theta)$ (dotted curves; the solid
lines denote the poloidal magnetic field/velocity stream lines; the
straight dashed line is the \Alfven surface) in a $10^8 r_o \times
10^8 r_o$ region for an ionizing luminosity $L =3.3 \times 10^{42}$,
erg~s$^{-1}$; this calculation of $\xi$ ignores the reduction of $L$
at large $r$ due to absorption. The values of the corresponding
quantities are noted on each dashed curve. The parameters are the
same as those of Figure~\ref{fig:wind}. [{\it See electronic edition
of the Journal for a color version of this figure.}] }
\label{fig:2d}
\end{figure}

With the density profile at hand, the wind ionization structure,
i.e. the local value of $\xi$, requires also an expression for the
X-ray luminosity. Since the models we present are global, i.e. they
cover a large number of decades in radius, it makes more sense that
we do not detach the ionizing luminosity from the global wind
dynamics. In order to do so we need a prescription that connects the
wind mass flux rate $\dot m_w$ and the ionizing luminosity $L$. The
simplest assumption is that $L \propto \dot m = \dot m_d = \dot
m_w$, i.e. to assume that the wind mass rate is proportional to the
accretion rate onto the black hole (we assume they are equal),
resulting in luminosity $L$ proportional $\dot m$. If this were
indeed true, then the ionization structure of the wind would be
independent of the accretion rate, since $\dot m$ would drop out of
the expression for the ionization parameter $\xi = L/nr^2$ [c.f.
eqn.~(\ref{eq:dnorm})]. However, this is contrary to observations
\citep[e.g.][]{Ueda03,Tueller08}, which indicate that the
probability of source obscuration decreases with increasing source
luminosity \citep[one should note, however, that obscuration may be
also due to gas in galactic mergers; e.g.][]{Hopkins05}. To
incorporate this fact, we provisionally adopt the prescription $L
\propto \dot m ^2$, even though any power near 2 or greater would
qualitatively suffice. We note that $L \propto \dot m^2$
prescription would be appropriate for flows such as those discussed
in \cite{NY95b} (provided that $\dot m \lsim \alpha^2$; $\alpha$ is
the usual disk viscosity parameter) with $L \propto \dot m$
otherwise; considering that the latter case is valid for only a
small range in $\dot m ~(0.1 \lsim \dot m \lsim 1)$ than the former,
we adopt the former, bearing in mind its limits of applicability.
Expressing the luminosity in terms of its Eddington value we can
write
\begin{equation}
L \simeq   \epsilon \dot m ^2 L_0 \hat{M} = 2\pi\, \epsilon
\dot m^2 \, \frac{r_S \, m_p c^3}{\sigma_{\rm T}} \ ,
\end{equation}
where $L_0 \simeq 1.28 \times 10^{38}$ erg~s$^{-1}$ is the Eddington
luminosity of a one solar mass an accreting object and $\epsilon$ is
the efficiency of conversion of mass into radiation for $\dot m =1$.
Given the ionizing luminosity $L$, and assuming the source to be
point-like, located at the coordinate origin, the local value of the
ionization parameter (ratio of ionizing flux per electron flux) is
\begin{eqnarray}
\xi(r,\theta) \equiv \frac{L}{n(r,\theta) r^2} \ . \label{eq:xi1}
\end{eqnarray}
Using the luminosity, density and radius scalings given above this
reads as
\begin{eqnarray}
\xi(r,\theta)
%&=&  \frac{\epsilon}{{\cal N}(theta) \eta_W} \;
%\frac{L_\odot \, \dot m}{n_0 \,r_S^2 \, x^{2q-1} }=
%\frac{\epsilon}{{\cal N}(\theta) \eta_W} \; \frac{L_0 \, \dot m
%\sigma_{\rm T}}{\tau_0 \,r_S     \, x^{2q-1}} \nonumber \\
&=& \frac{4 \pi \, \epsilon}{  {\cal N}(\theta) \eta_W} \; \frac{m_p
c^3\, \dot m}{x^{2q-1}} \simeq  \frac{\epsilon}{{\cal N}(\theta)
\eta_W} \; \frac{3 \times 10^8 \, \dot m}{ x^{2q-1}} \ .
\label{eq:xi1}
\end{eqnarray}
Equation~(\ref{eq:xi1}) thus provides the explicit dependence of the
ionization parameter of the corresponding wind as a function of the
accretion rate $\dot m$, the (normalized) distance from the source
$x$ and the LOS angle $\theta$ for steady-state accretion at a rate
$\dot m$ into the black hole and actual mass outflow rate (at radius
$x \simeq 1$) $\eta_W \dot m$ in the wind. Figure~\ref{fig:2d} shows
poloidal color maps of (a) the wind density $n(r,\theta)$ and (b)
the ionization parameter $\xi(r,\theta)$ for the parameters used to
produce Figure~\ref{fig:wind} along with a number of contour curves
(dotted curves with values) and the flow streamlines (solid curves).
The diagonal dashed straight lines denote the self-similar surface
of the \Alfven point. Here, we have assumed $M_{BH} =
10^{6}M_{\odot}$ and $L =3.3 \times 10^{42}$ erg~s$^{-1}$, which for
$\epsilon \simeq 0.2$ implies $\dot m \simeq 0.1$. As seen, we
obtain a non-recollimating, non-oscillating wind which
asymptotically behaves as $r \propto (\cot \theta)^{1/\zeta} \sec
\theta$ where $\zeta=\textmd{const.} \simeq 1/2$ as $r \rightarrow
\infty$ [or in cylindrical coordinates $(Z,R)$ $R \propto
Z^{1-\zeta}$; see CL94]. We would like to stress that the field line
geometry depends on the wind's conserved quantities in the model. As
a consequence the obtained wind properties (shown here in
Figs.~\ref{fig:wind} to \ref{fig:2d}) can change quite a bit.

%$R \propto Z^{1-\epsilon}$ where
%$\epsilon=\textmd{const.} \simeq 1/2$ as the cylindrical
%coordinates ($Z,R) \rightarrow \infty$ (see CL94).

%
%-----------------------------------------------Place Figure~3
%-------------------------------------------------------------
%

In the context of our accretion-disk wind scenario the outflows are
necessarily more opaque near the launching disk surface ($\theta =
30\deg$) at smaller radius as expressed in equation~(\ref{eq:n}).
Density profile along a magnetic field follows exactly the one shown
in Figure~\ref{fig:density}a. There are several points to note about
equation~(\ref{eq:xi1}) and Figure~\ref{fig:2d}b: ($i$) If distances
are scaled by the Schwarzschild radius, $r_S$, and the accretion
rate by its Eddington value, $\dot M_{\rm Edd}$, the ionization
parameter, $\xi(r, \theta)$, depends primarily on the normalized
accretion rate $\dot m$ and ${\cal N}(\theta)$; as a result, our
analysis and the corresponding ionization structure of the wind, are
applicable to accreting black holes of any size. Note that, had one
assumed $L \propto \dot m$, the expression for $\xi$ would be
independent even of $\dot m$; we believe that this would be in
disagreement with observation; $({ii})$ The ionization parameter
$\xi$ is much higher near the central engine (e.g. within a few
hundreds to thousands \sw radii) with the wind highly ionized in
this region and for a given $r$ it drops precipitously with
$\theta$; $({iii})$  For a given $\theta$, the value of $\xi$ drops
with distance (with $\xi \propto 1/r$ for our favorite value of $q
\simeq 1$); ($iv$) $\xi$ increases linearly with increasing
accretion rate $\dot m$, implying decreasing source obscuration for
high values of $\dot m$ as seen in equation~(\ref{eq:column1}), in
agreement with observation \citep{Tueller08}; ($v$) The wind is
fully ionized in regions sufficiently close to the central engine
thus no contribution to the absorption lines in the spectrum.

One should note that, as written, the expression of equation~
(\ref{eq:xi1}) is a local one, i.e. it does not take into account
neither the absorption of the X-ray radiation in its propagation
along the LOS wind nor the corresponding re-emission, both of which
would modify the local ionizing spectrum, and thus also the global
ionization structure shown in Figure~\ref{fig:2d}b here. These
effects are small at small values of the radius ($\log x \lsim 3$
and for relatively hard spectra $\alpha \simeq 1-2$), where the
ionization of the gas is high and its absorption depth small at all
energies; they become more significant at larger radii and they are
taken into detailed consideration in our model calculations as
discussed below.

It should also be cautioned at this point, that while $\xi(r,
\theta)$ is the primary parameter that determines the ionization
state of the X-ray illuminated gas, the latter depends also on the
spectrum of the incident radiation, as the resulting electron
temperature is roughly equal to the mean photon energy of the
ionizing radiation. In the calculations presented herein we assume a
single power-law with an energy spectrum $L_\nu \propto
\nu^{-\alpha}$ between 13.6 eV and 13.6 keV and $\alpha = 1.5$
as a fiducial model \citep[e.g.][]{Sim08}. More realistic incident
spectral distributions \citep[those used, e.g.,
in][]{Everett05,Sim05,SD07} can be easily accommodated within
\verb"XSTAR" likely leading to results that are not substantially
different from those of the next sections, although we defer such
studies to future publications.

%Figure~\ref{fig:xi} shows a poloidal plane color map of the
%ionization parameter $\xi(r,\theta)$, as given by equation~
%(\ref{eq:xi1}) for the parameters used to produce
%Figure~\ref{fig:wind} along with a number of equal $\xi(r, \theta)$
%contour lines (dotted curves with values) and the flow streamlines
%(solid curves). A gray straight line denotes the self-similar
%position of the \Alfven point. As seen, we obtain a
%non-recollimating, non-oscillating wind which asymptotically behaves
%as $R \propto Z^{1-\epsilon}$ where $\epsilon=\textmd{const} \simeq
%1/2$ as the cylindrical coordinates ($Z,R) \rightarrow \infty$ (see
%CL94).

\subsubsection{Radiative Transfer in the Wind}
As noted above, equation~(\ref{eq:xi1}) is strictly valid when the
attenuation of the radiation field by the wind itself is small; it
is used in the computation of the ionization state only for the
innermost, highly ionized regions of the wind, while at larger radii
we use the absorbed radiation field is employing \verb"XSTAR" for
the computation of the relevant opacities; the method is described
below and it is very similar to the recent work of, e.g.,
\citet{SD07}, \citet{DKP08} and \citet{SD09}.
%
%We further define dimensionless variables as $n_{10} \equiv n_0/10^{10}$,
%$x \equiv r/ r_s$, and $x_0 \equiv r_0 r_s$ where $n_{10}$ is the wind
%density in units of $10^{10}$ cm$^{-3}$ and $r_s \backsimeq 3 \times
%10^{11} M_6$ cm is the \sw radius scaled by .
%
Since at present we are interested mainly in the absorption
properties of the above configuration, whose Thomson opacity is
generally small, we treat the gas along the LOS independently as a
collection of radially discrete slabs of constant (local) density,
and the total column depends only on the LOS angle $\theta$ in
accordance with Figure~\ref{fig:density}a. At large distances
where the value of the ionization parameter drops and He and H
opacities become important, we let \verb"XSTAR" subdivide each
zone as necessary for the proper treatment of the radiative
transfer; since in this case the absorption mean-free path at the
characteristic frequencies is much smaller than the local radius,
the plane parallel geometry employed by \verb"XSTAR" for this
purpose is adequate.

With the ionizing luminosity $L$ and a spectrum specified at $x=1$,
the entire LOS wind with the radial length scale of $\log (\Delta x)
\simeq 10$ is divided up into a number of (plane parallel)
slabs/zones (typically $\sim 40$) of logarithmically-equal thickness
in radius; i.e. $\Delta x/x = \textmd{const}$.
%In subsequent sections our discussion on the model spectra are thus based on the
%calculations with these parameter sets.
Then, using \verb"XSTAR" we calculate the absorption and emission
coefficients within, say, the $(i+1)$-th zone, using as input the
(properly normalized) output spectrum of the $i$-th zone, $L_i$. We
thus obtain the forwardly emitted spectrum in emission lines
$L_{i+1}^{\rm (line)}$ and in continuum $L_{i+1}^{\rm (cont)}$ by
computing the free-free, free-bound and bound-bound atomic
transitions within the $(i+1)$-th zone (see \verb"XSTAR" manual),
while we ignore for this calculation the part of the backward
directed flux. We also compute the fraction of the radiation flux of
the $i$-th zone, $L_i$, that is transmitted through the $(i+1)$-th
zone, $L^{\rm (tr)}_{i+1}$, given by
\begin{eqnarray}
L^{\rm (tr)}_{i+1} =  L_i e^{-\tau_{(i+1)}} \ , \label{eq:trans}
\end{eqnarray}
where $\tau_{i+1}(N_{H,i})$ is the optical depth (including
continuum and line photoabsorption as well as Thomson scattering) of
the $(i+1)$-th zone, with $N_{H,i}$ being the total hydrogen
equivalent column of this slab.
Finally, we produce the (properly normalized) spectrum of the
$(i+1)$-th zone, by adding the forwardly emitted and transmitted
spectra, i.e.
\begin{eqnarray}
L_{i+1} =  L^{\rm (tr)}_{i+1} + L^{\rm (cont)}_{i+1} +
L^{\rm (line)}_{i+1} \ , \label{eq:emission}
\end{eqnarray}
with the process being iterated with $L_{i+1}$ in place of $L_i$, to
compute $L_{i+2}$ and so on.

\subsubsection{Modeling Absorption Line Profiles}
The above procedure can be applied to a specific atomic transition
to produce the resulting line profile and an absorption feature
since this is the dominant process along the observer's LOS to the
point-like ionizing source. The corresponding emission is
distributed isotropically and makes a negligible contribution to
specific intensity at the particular frequency. The radiation
transfer of a specific feature is dominated by the optical depth
of the wind at the specific frequency $\tau(\nu)$, which is
expressed by
\begin{eqnarray}
\tau(\nu) = \sigma(\nu) N_H(\nu)   \ , \label{eq:tau1}
\end{eqnarray}
with the line photoabsorption cross-section
\begin{eqnarray}
\sigma=0.01495 (f_{ij} / \Delta \nu_D) H(a,u) \ , \label{eq:sigma}
\end{eqnarray}
where $f_{ij}$ is the oscillator strength of the transition between
the $i-$th and $j-$th levels of an ionic species and $\Delta \nu_D$
is the Doppler broadening factor estimated by $\Delta \nu_D \approx
(\Delta v_{\rm turb}/ c) \nu_0$ relative to the centroid
(rest-frame) frequency $\nu_0$ with $v_{\rm turb}$ being either
thermal or turbulent velocity of the medium, assuming that the
overall absorption profiles are well approximated locally by the
Voigt function $H(a,u)$ \citep[e.g.][]{Mihalas78,Kotani00,Hanke09}.
This represents a line profile whose centroid is dominated by the
Doppler broadening while the wings are characterized by the
damping/Lorentzian profile and defined by
\begin{eqnarray}
H(a,u) \equiv \frac{a}{\pi} \int_{-\infty}^\infty \frac{e^{-y^2}
dy}{(u-y)^2+a^2}   \ .\label{eq:voigt}
\end{eqnarray}
In the above expression, we use $a \equiv \Gamma / (4 \pi \Delta
\nu_D)$ where $\Gamma$ is the Einstein coefficient and $u \equiv (\nu
- \nu_0)/\Delta \nu_D$ is the dimensionless frequency spread about
the transition frequency. Irrespective of their thermal or turbulent
velocities, the winds we consider provide a well defined velocity
shear between two adjacent slabs, whose effect on the radiative
transfer of a line photon is very similar to that of the turbulent
velocity. To rid the situation of as much unwanted uncertainty as
possible we then consider the thermal and turbulent broadenings to
be negligible and we consider only the shear broadening, so that
\begin{eqnarray}
\Delta \nu_D \approx (\Delta v_{\rm turb}/ c) \nu_0 = (\Delta v_{\rm
sh}/c) \nu_0 \ ,
\end{eqnarray}
with $\Delta v_{\rm sh}$ being the LOS velocity shear. In addition
to the broadening, the velocity of the absorbing medium shifts also
the transition frequency blueward as $\nu'_0 = \nu_0/[1 - v_{\rm
los}(x, \theta)/c]$ where $v_{\rm los}(x, \theta) = v_r(x, \theta) $
is the flow velocity along LOS and $\nu'_0$ is the overall observed
frequency. With these considerations the expression for the
dimensionless frequency spread about the specific transition reads
\begin{eqnarray}
u(x, \theta) = \frac{\nu/\nu_0 - 1/[1-v_{\rm los}(x)/c]}{\Delta
v_{\rm sh}/c} \ .
\end{eqnarray}
%
%where
%
%\begin{eqnarray}
%\beta_{\rm los}(r,\theta)  \equiv v_{p} \sin \left[\theta+\cot^{-1}
%(R')\right] \ ,
%\label{eq:v_los}
%\end{eqnarray}
%
%is defined locally by the poloidal wind velocity  $v_p(r, \theta)$
%and the streamline geometry $R(Z)$ with $R' = dR/dZ$.

\subsubsection{Modeling the Absorption Measure Distribution (AMD)}
In a wind model with well-defined density as a function of radius
along the LOS, it is natural to use the value of $\xi$, the
parameter that determines the presence (or not) of a given ion, as a
proxy for the radius $r$ along the wind in order to study the
structure of its ionization. Then the hydrogen equivalent column of
specific ions, $N_H$, as a function of $\xi$  can be used to
determine the wind density distribution as a function of radius.
This has been precisely the approach of HBK07, who defined the AMD
as the hydrogen equivalent ion column density per logarithmic
ionization parameter interval; in view of our specific wind models,
this has the following dependences [Eqns.~(\ref{eq:column1}) and (\ref{eq:xi1})]
\begin{eqnarray}
{\rm{AMD}}(q;\xi) \equiv \frac{d \left[N_{\rm H}(r,\theta)\right]}{d
\left[\log \xi(r,\theta) \right]} \propto r^{2(q-1)} \propto
\xi^{\frac{2(q-1)}{1-2q}} \ . \label{eq:amd}
\end{eqnarray}
Therefore, depending on the specific run of the wind density with radius
%radial dependence of the models winds
(via $q$), the AMD is constant, i.e., independent of the radius $r$
or ionization parameter of the plasma $\xi$ when $q=1$, while it
scales as ${\rm{AMD}} \propto r^{-1/2} \propto \xi$, i.e.,
monotonically increasing with ionization $\xi$ for the BP82 models
($q=3/4$).

\section{Results}

In this section we discuss in detail the implementation of the
procedures described above and the ensuing results concerning the
properties of individual atomic transitions as well as their
dependence on the parameters of the problem. In particular we focus
on specific ionic charge states that are important in that they can
serve as spectroscopic diagnostics of the ionization state of the
wind medium. Because of the specific, continuous decrease of the MHD
wind density inherent to the accretion-origin wind scenario, our
models span a wide range of ionization parameter space and hence a
large number of the corresponding atomic transitions (in
absorption). Their global properties are encapsulated in the AMD,
namely the distribution of (locally) absorbing column per
logarithmic ionization parameter interval, a quantity originally
introduced by HBK07, which can and has been compiled for a number of
AGNs using comprehensive fits to the ensemble of absorption features
in their entire X-ray spectrum \citep[][]{Behar09}. Furthermore,
given that to each ionization state of the plasma corresponds also a
specific wind velocity, there exists a correlation between the
presence of a specific atomic transition and the corresponding
plasma velocity. Employing this generic property, we demonstrate a
progressive absorption of the incident X-ray by matter outflowing at
the different velocities (as discussed in \S 2.2), to produce the
eventual absorption line profile of specific transitions. Finally,
because of the strong dependence of the wind column density along
the LOS angle $\theta$, as discussed in \S 2.1, we consider two
representative cases for comparison [employing always the $q=1$
field geometry and the parameters used in producing
Figs~\ref{fig:wind}-\ref{fig:2d}], namely $\theta=30\degr$ (high
latitude LOS) and $60\degr$ (low latitude LOS).

Considering the underlying complexity of the observed broad-band
SEDs of AGNs we have decided to adopt the simple prescription of a
single power-law form for the ionizing flux discussed above  and use
different values of its slope $\alpha$ to simulate the diversity of
observed AGN SEDs. This approach, while it should be viewed with
caution vis-\'a-vis the detailed properties of the wind ionization,
we believe that it is sufficient in capturing the trends of the
winds' ionization properties with changes in the overall AGN SED.
The photoionization calculations we present include the most
important elements (H, He, Ca, Na, O, Ne, Mg, Si and Fe) with
abundances set to the solar values \citep[][]{Grevesse96}. Based on
this set-up the major characteristics of our hydromagnetic wind are
listed in Table~\ref{tab:tbl-1} which we shall discuss in subsequent
sections.

%
%-------------------------------------------------------------- Table~1
\begin{deluxetable}{c||c|l}
\tabletypesize{\scriptsize}
\tablecaption{Characteristic Wind
Properties of Our Models. \label{tab:tbl-1}} \tablewidth{0pt}
\tablehead{{\it Parameter}~$^a$ & {\it Fiducial Value}
($\theta=30\degr/60\degr$) & {\it Physical Significance}  }
\startdata
$\hat{M}$ & $10^6$ & Mass of the central black hole in units of $\Msun$ \\
$\alpha$ & $-1.5$ & Power-law index of the incident spectrum (in energy) \\
$L_X$ & $3.3 \times 10^{42}$ erg~s$^{-1}$ & Incident X-ray luminosity   \\
$r_o$ & $\sim r_s$ & Radius of the innermost foot point of the wind \\
$q$ & 1 & Scaling of magnetic fields \\
$\dot{m}$ & 0.1 & Dimensionless (conserved) mass-accretion rate \\
$\eta_W$ & 0.5 & Outflow (wind) rate relative to mass-accretion rate
$\dot{m}$ \\
$\epsilon$ & 0.2 & Radiative efficiency \\
$x_{\rm in}$ & $6.7 / 2.0 $  & Dimensionless LOS Inner radius of the
wind  \\
%$x_{out}$ & $1.6 \times 10^{21} / 1.4 \times 10^{21}$ cm & LOS Outer
%radius of the wind \\
$n(x_{\rm in})$ & $1.9 \times 10^9 / 4.5 \times 10^{10}$ cm$^{-3}$
&
Density of the wind at $x=x_{\rm in}$ \\
$\Delta x$ & $\sim 10^{10}$ & LOS length scale of wind \\
%$n(r_{out})$ & $0.1 / 1.1 $ cm$^{-3}$ & Density of the wind at
%$r=r_{out}$ \\
%$v_{\rm los}(x_{\rm in})$ & $\sim 237,000 / 117,000$ km~s$^{-1}$ &
%LOS
%wind velocity at $x=x_{\rm in}$ \\
$\log \xi(x_{\rm in})$ & 8.62 / 8.3 & Ionization parameter at
$x=x_{\rm in}$ \\
$v_{\rm los}$(\fexvii) & $\sim 100-300$ km~s$^{-1}$ &
Characteristic LOS velocity of \fexvii \\
$\log \xi$(\fexvii) & $\sim 2.2-3$ & Ionization parameter of \fexvii \\
$\log x$(\fexvii) & $\sim 6-7 / 5-6$ & LOS distance of \fexvii \\
$v_{\rm los}$(\fexxv) & $\sim 2,000 - 4,000 / 1,000 - 3,000$
km~s$^{-1}$ & Characteristic LOS velocity of \fexxv \\
$\log \xi$(\fexxv) & $\sim 4-5$ & Ionization parameter of \fexxv \\
$\log x$(\fexxv) & $\sim 4-5 / 3-4$ & LOS distance of \fexxv \\
$v_{\rm los}$(\ovii) & $\sim 40 - 150$ km~s$^{-1}$ & Characteristic
LOS
velocity of \ovii \\
$v_{\rm los}$(\oviii) & $\sim 100 - 600 / 100-300$ km~s$^{-1}$ &
Characteristic LOS velocity of \oviii \\
$\Delta N_H$ & $\sim 2.6 \times 10^{21} / 1.8 \times 10^{22}$
cm$^{-2}$ &
Local wind column density per slab \\
$N_H$ & $\sim 3.9 \times 10^{22} / 2.5 \times 10^{23}$ cm$^{-2}$ &
Integrated wind column density over $-1 \lesssim  \log \xi \lesssim 4$ \\
%$\beta_{los}(x_{out})$ & 3.7 / 4.0 km~sec$^{-1}$ &  LOS wind velocity
%at $r=r_{out}$ \\
\enddata
\vspace{0.05in} $^a$ See the text for the template parameter values
in detail.
\end{deluxetable}
%---------------------------------------------------------------
%

\subsection{Absorption Measure Distribution (AMD) }

\begin{figure}[t]% ------------------------------------- Figure~4
\centering
\begin{tabular}{cc}
\epsfig{file=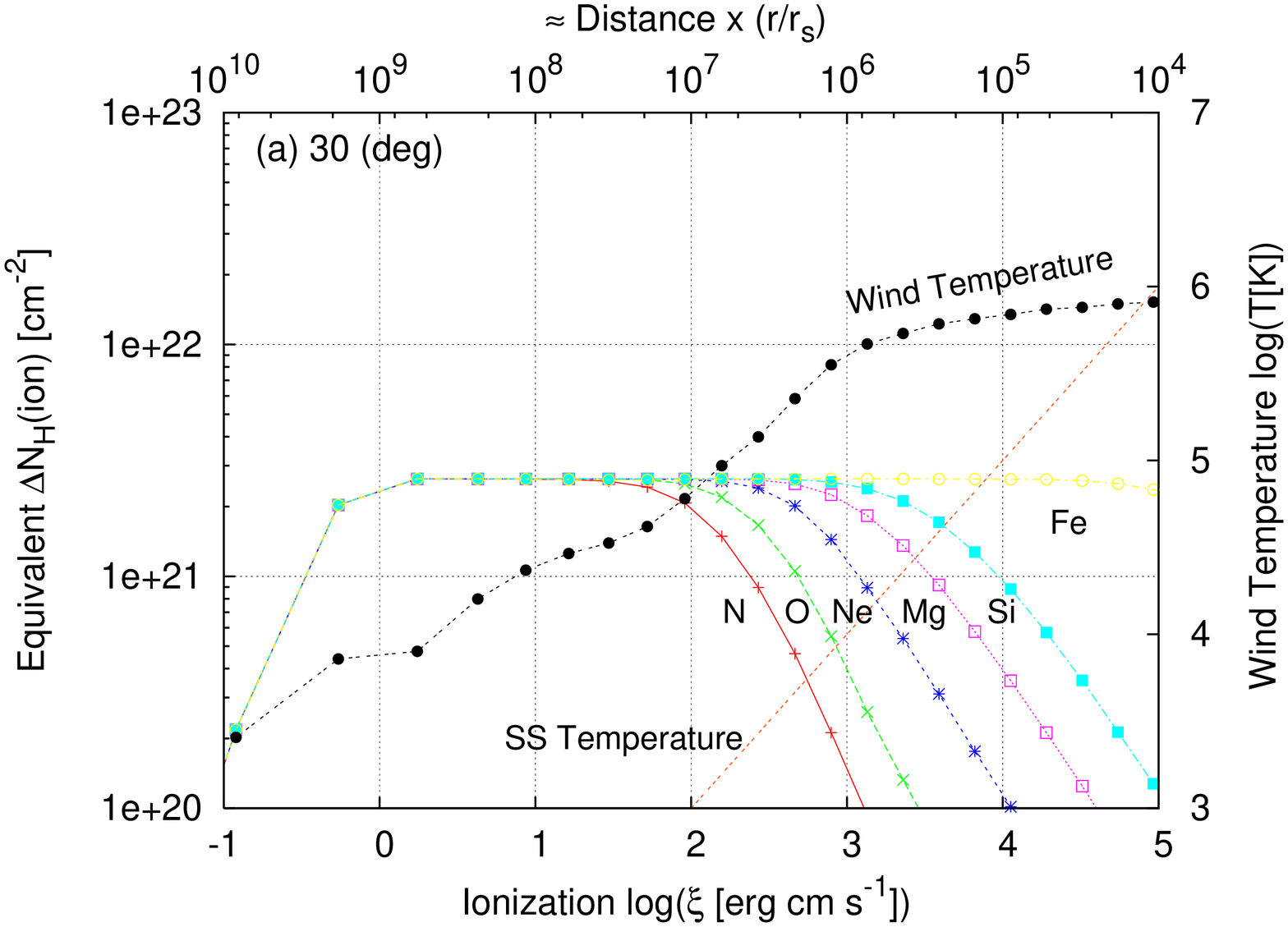,angle=-90,width=0.45\linewidth,clip=""} &
\epsfig{file=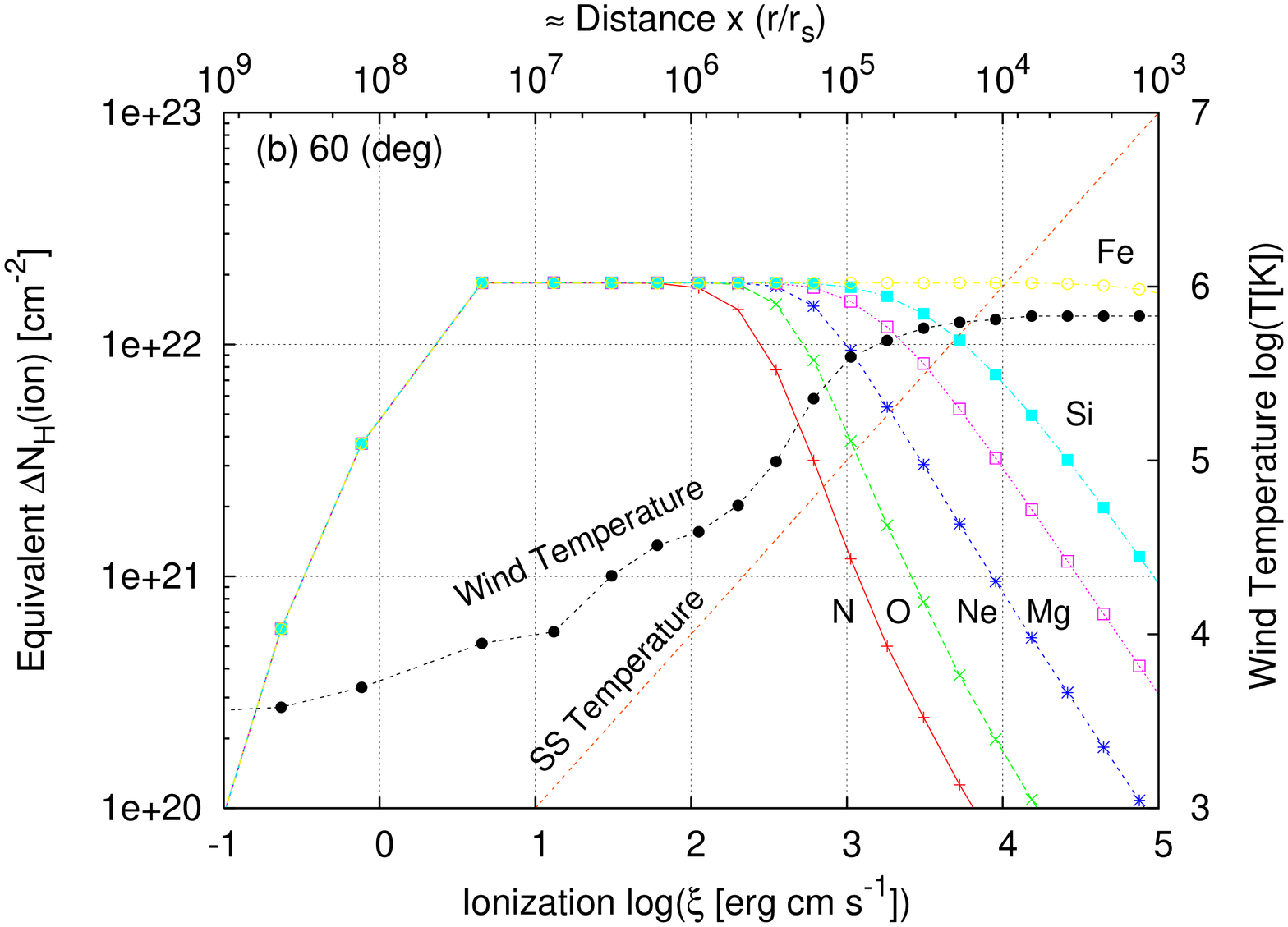,angle=-90,width=0.45\linewidth,clip=""}
\end{tabular}
\caption{Simulated hydrogen-equivalent (local) column density (per slab) $\Delta
N_H$(ion) for major ionic elements (N, O, Ne, Mg, Si and
Fe) and wind temperature $T$ (right axis) in the fiducial model for (a)
$30\degr$ and (b) $60\degr$. Analytic self-similar (SS) wind temperature is also shown. Corresponding LOS (normalized) distance
$x(\equiv r/r_S)$ is also shown in the upper axis. The parameters
are the same as in Figure~\ref{fig:wind}. Rapid decrease of $\Delta N_H$ in the highest (lowest) end of $\xi$ is due to fully stripped ions (neutral ions). [{\it See the electronic
edition of the Journal for a color version of this figure.}]
\label{fig:amd}}
\end{figure}

The unique characteristic of any wind model ionized by a
point-source located at its origin is the radial distribution of
density and therefore the corresponding distribution of the
ionization states of the different elements seen in absorption in
the source continuum. A particular wind model is therefore
characterized by the distribution of the ionization states of the
different elements, their velocities and the distribution of their
column densities or their (local) equivalent hydrogen column $N_H$;
these determine the run of the wind's density and velocity with the
radius $r$. For example, in the accelerating phase of a spherical
wind, both the ionization parameter and the velocity increase with
distance and ``freeze" once it has achieved its terminal velocity,
while the column density of ions decreases faster than $1/r$ if an
ion is present in the wind accelerating stage, leveling off to $N_H
\propto 1/r$ once terminal velocity is achieved.

The characteristic properties of the winds presented in \S 2 are
drastically different from those of spherical winds. In particular:
($i$) The ionization parameter $\xi$ and the corresponding LOS
velocity $v_{\rm los}$ decrease with distance as $r^{-1}$ and
$r^{-1/2}$ respectively to encompass a large number of ionization
states of many elements at different velocities. ($ii$) The specific
column density per decade of radius translates for any element to a
(local) equivalent hydrogen column $\Delta N_H$ that is independent
of the radius $r$.

For a given wind model and ionizing spectrum, one can produce the
corresponding AMD profiles, i.e. the global distribution of its
ionization properties which can then be compared to observations. We
do so in this section and compare the results to the recent AMD
analysis of IRAS~13349+2438 observations (HBK07). The results of our
calculations are shown in Figure~\ref{fig:amd} where we present the
AMD for the $q=1$ wind models, with the $\Delta N_H$ plotted as a
function of $\log \xi$ for $\theta=30\degr$ in (a) and $60\degr$ in
(b). Also superimposed is the plasma (gas) temperature $\log T$
computed by \verb"XSTAR" under local thermal equilibrium conditions
along with the analytic self-similar (SS) temperature ($\propto
1/r$). The ionized wind temperature decreases monotonically from $ T
\sim 10^6$ K down to $T \sim 10^{3.5}$ K as $\xi$ decreases from
$\log \xi \simeq 5$ to  $\log \xi \simeq -1$ a range relevant to the
observable ionic states of the major elements. For reference, the
corresponding (dimensionless) LOS distance $x \equiv r/r_S$ from the
central engine is also shown in the upper x-axis where in this run
$r_S \cong 3 \times 10^{11}$ cm. As seen, the local column densities
of the irradiated ions (shown here are the six major elements: N, O,
Ne, Mg, Si and Fe) are found to be all distributed at constant value
of $\Delta N_H$ over many decades of $\xi$ (and therefore a similar
range in LOS distance too), namely $\Delta N_H \sim 2.6 \times
10^{21}$ cm$^{-2}$ for $30\degr$ and $\sim 1.8 \times 10^{22}$
cm$^{-2}$ for $60\degr$, reflecting the different wind density at
these different LOS (see Fig.~\ref{fig:density}a). This results in
an integrated column of $N_H \sim 3.9 \times 10^{22}$ cm$^{-2}$ for
$30\degr$ and $\sim 2.5 \times 10^{23}$ cm$^{-2}$ for $60\degr$ over
$-1 \lesssim \log \xi \lesssim 4$. The constant value of AMD is a
unique characteristic of the $q=1$ models with the value
corresponding to case (a) ($\theta =30\degr$) in good agreement with
the observed AMD of IRAS~13349+2438 analyzed by HBK07.
%
%The difference is simply due to the fact that the wind becomes more
%opaque toward the disk (or equatorial bulk flow). .
%

As shown here, the computed temperature has a radial dependence
(gradient) close to the desired $1/x$ one for $x \gsim 10^5$, but it
is constant for $x \lsim 10^5$. However, because the thermal
pressure is smaller than the wind ram pressure over the entire
domain, the thermal pressure effects on the wind density profiles
are not significant. We have also compared the heating/cooling time
scales of \verb"XSTAR" and were found to be shorter than the local
dynamical time scale $r/v$, indicating that the temperature
distributions of Figure~\ref{fig:amd} are indeed correct, even
though not dynamically important. We have obtained by another
calculation a wind solution similar to Figure~\ref{fig:amd} in the
cold-flow limit ($K=0$).

%High quality X-ray absorption data were also obtained for the GBHC
%GRO~J1655-40 \citep{Miller06} and were used to argue for magnetic
%driving of the wind in this system too. These spectra are
%distinguished from those of AGNs by the prominent absence of low
%ionization state ions. This is to be expected given that the
%presence of the companion star limits the extent of the disk to
%roughly half the distance between the two objects or $r \simeq
%10^{12}$ cm. Given that the Schwarzschild radius of the compact
%object is $r \simeq 10^6$ cm the entire disk size covers only a range
%of $x \sim 10^6$ in radius; considering (based on Fig.~\ref{fig:amd})
%that in the inner region of $r \sim 1000-3000 r_S \sim 10^9$ cm the
%elements are fully ionized that leaves ions over only a factor of
%1000 in $\xi$, in rough agreement with observation.

It should be noted that the decrease of $\Delta N_H$ at high values
of $\xi$ is due to the fact that \verb"XSTAR" by default does not
provide the column density of fully ionized (bare) ions. This
apparent decrease is just due to the fact that all considered
elements become progressively (completely) ionized at high $\xi$
values. At the other end of the plot, i.e. for small values of
$\xi$, $\Delta N_H$ appears to decrease too because the wind becomes
too cold (gas temperature of $T \lesssim 10^{3.5}$ K) to participate
in radiative processes, i.e. each species turns neutral, an
ionization state also not included in the calculation of the
hydrogen equivalent column. The calculated AMD is otherwise
constant, as expected. The fact that the corresponding $\Delta N_H$
is the same in this regime irrespective of the elements used in the
computation is a testimony that our models implement \verb"XSTAR"
correctly, so that the sum of the abundances of all ions of any
element, when corrected for the elemental abundance, yields the same
value of $\Delta N_H$. We also see that the (relatively) lighter
elements (say, oxygen) become fully ionized at lower $\xi$ than
heavier ones (say, iron), also as expected. In essence, the overall
trend of the model AMD in both LOS cases (a) and (b) appears to be
very similar to one another (i.e. constant value of $\Delta N_H$
independent of ionization state) except for the fact that the local
column $\Delta N_H$ for $30\degr$ in (a) is about a factor of $\sim
10$ lower than that for $60\degr$ in (b) because of the angular
dependence of the wind density function ${\cal N}(\theta)$ (also see
Fig.~\ref{fig:density}a).

\begin{figure}[t]% ------------------------------------- Figure~5
\centering
\begin{tabular}{cc}
\epsfig{file=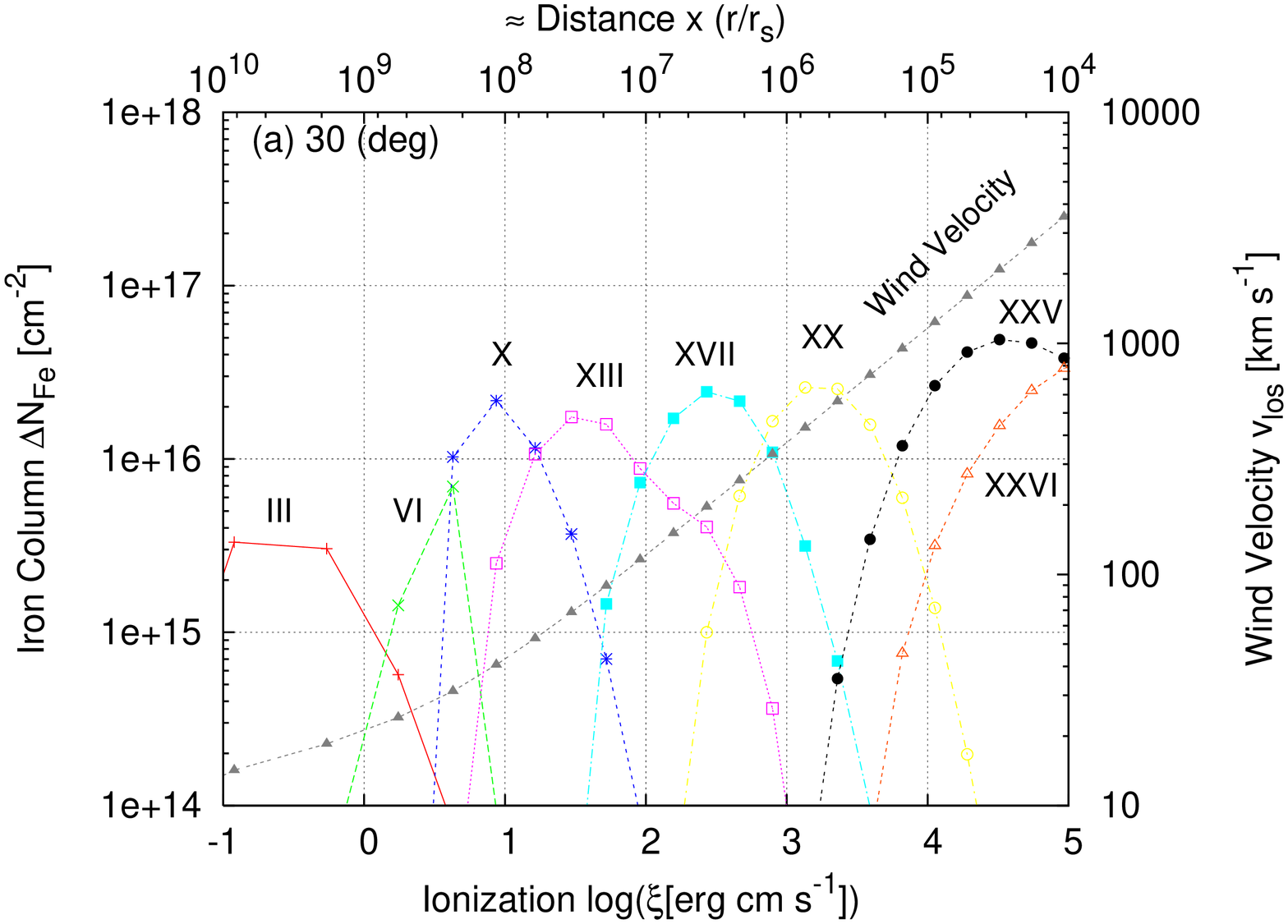,angle=-90,width=0.45\linewidth,clip=""} &
\epsfig{file=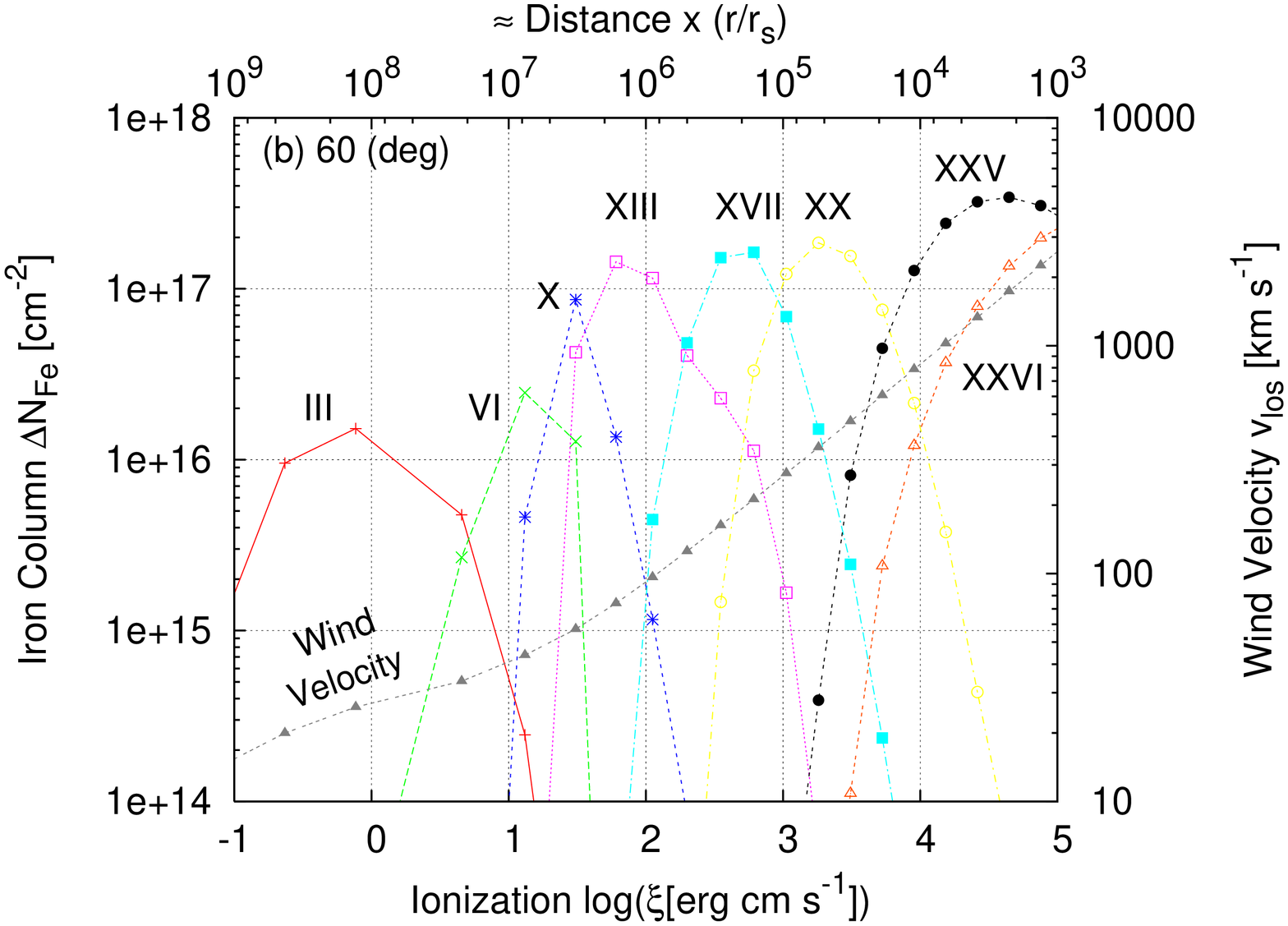,angle=-90,width=0.45\linewidth,clip=""}
\end{tabular}
\caption{Simulated local iron column density $\Delta N_{\rm Fe}$ distribution
for various iron charge levels (\feiii, \fevi, \fex, \fexiii,
\fexvii, \fexx, \fexxv, and \fexxvi) and LOS velocities $v_{\rm los}$ (right axis)
for (a) $30\degr$ and (b) $60\degr$ with the same fiducial model
parameters as in Figure~\ref{fig:amd}. [{\it See the electronic
edition of the Journal for a color version of this figure.}]
\label{fig:fe}}
\end{figure}

It is also of interest to show the differential form of
Figure~\ref{fig:amd}, i.e. the columns of specific charge states of
a given ionic element $\Delta N_{\rm ion}$. This is shown in
Figure~\ref{fig:fe} for iron again for two different LOS. This
figure shows the emergence of various ionization states of iron
(shown here are \feiii, \fevi, \fex, \fexiii, \fexvii, \fexx, \fexxv
~and \fexxvi) along with the corresponding LOS velocity $v_{\rm
los}$ for $\theta=30\degr$ in (a) and $60\degr$ in (b) corresponding
to Figure~\ref{fig:amd}. As discussed above, the higher ionization
states are present at the inner sections of the wind (smaller $x$),
being replaced by lower ionization ions at larger radii with their
columns given in absolute values $\Delta N_{\rm Fe}$ instead of the
hydrogen-equivalent values $\Delta N_H$. One should note that the maximum
value (peak) of $\Delta N_{\rm Fe}$ for a specific charge state is
approximately constant until one gets to the lowest ionization
states, located at the largest distances \citep[see, e.g.,][for a
similar transition]{Murray98} because the fraction of neutral iron
relative to the total iron starts increasing as mentioned earlier
for Figure~\ref{fig:amd}.
%
%Their maximum value is smaller because the X-ray flux is lower than
%that estimated from the simple geometric dilution, due to its
%absorption by matter at the inner radii.
%
Of interest is also the LOS wind velocity $v_{\rm los}$ that
corresponds to a given value of $\xi$. While $v_{\rm los}$ decreases
like $\propto \xi^{1/2}$ at high values of $\xi$ (as $\xi \propto
r^{-1}$ and $v_{\rm los} \propto r^{-1/2}$), it deviates from this
dependence at the largest radii because of the additional reduction
of the X-ray flux due to absorption. The velocity at which each ion
has its maximum column and the range in $v_{\rm los}$ over which it
has substantial column are of interest because they determine the
absorption line profile of the specific ion. Thus, in the case of
(a) $\theta=30\degr$, higher charge states such as \fexxv~ are found
to be blueshifted by a relatively high outflow velocity, $v_{\rm
los} \sim 2,000-4,000$ km~s$^{-1}$ at $\log \xi \sim 4-5$ ($\log x
\sim 3-4$) while lower charge states such as \fexvii~ are
blueshifted by only $v_{\rm los} \sim 100-300$ km~s$^{-1}$ at $\log
\xi \sim 2.2-3$ ($\log x \sim 6-7$). We emphasize that the latter
(\fexvii) line velocity and column density are also in good
agreement with those of IRAS~13349+2438 (HBK07). For $\theta =
60\degr$ in comparison, the column density increases $\sim 10$ times
but the corresponding line velocity and ionization stage remains
roughly the same (but at smaller distances) because both the
poloidal velocity and its projection to the LOS are smaller at lower
latitudes (see Fig.~\ref{fig:density}b).

\begin{figure}[t]% ------------------------------------- Figure~6
\centering
\begin{tabular}{cc}
\epsfig{file=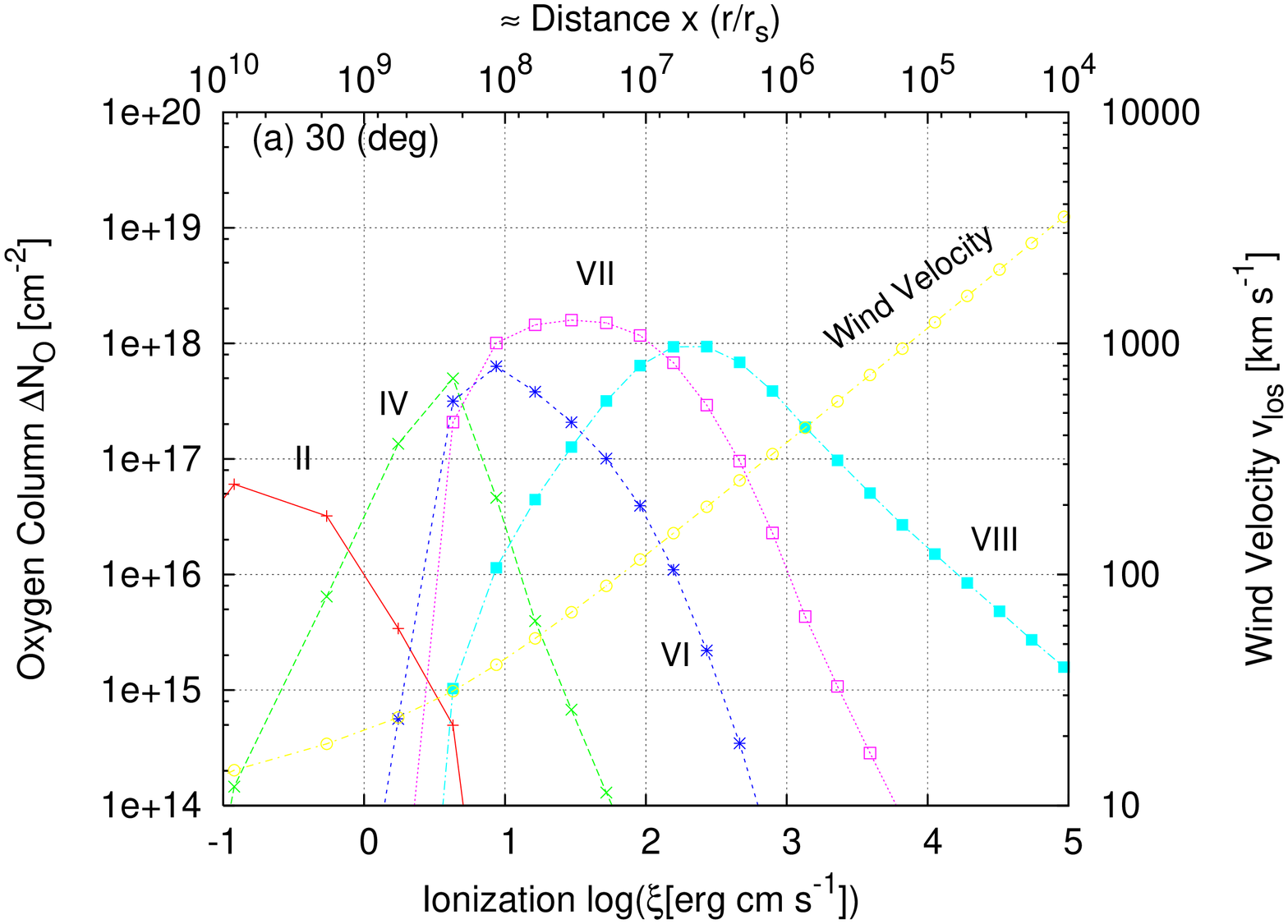,angle=-90,width=0.45\linewidth,clip=""} &
\epsfig{file=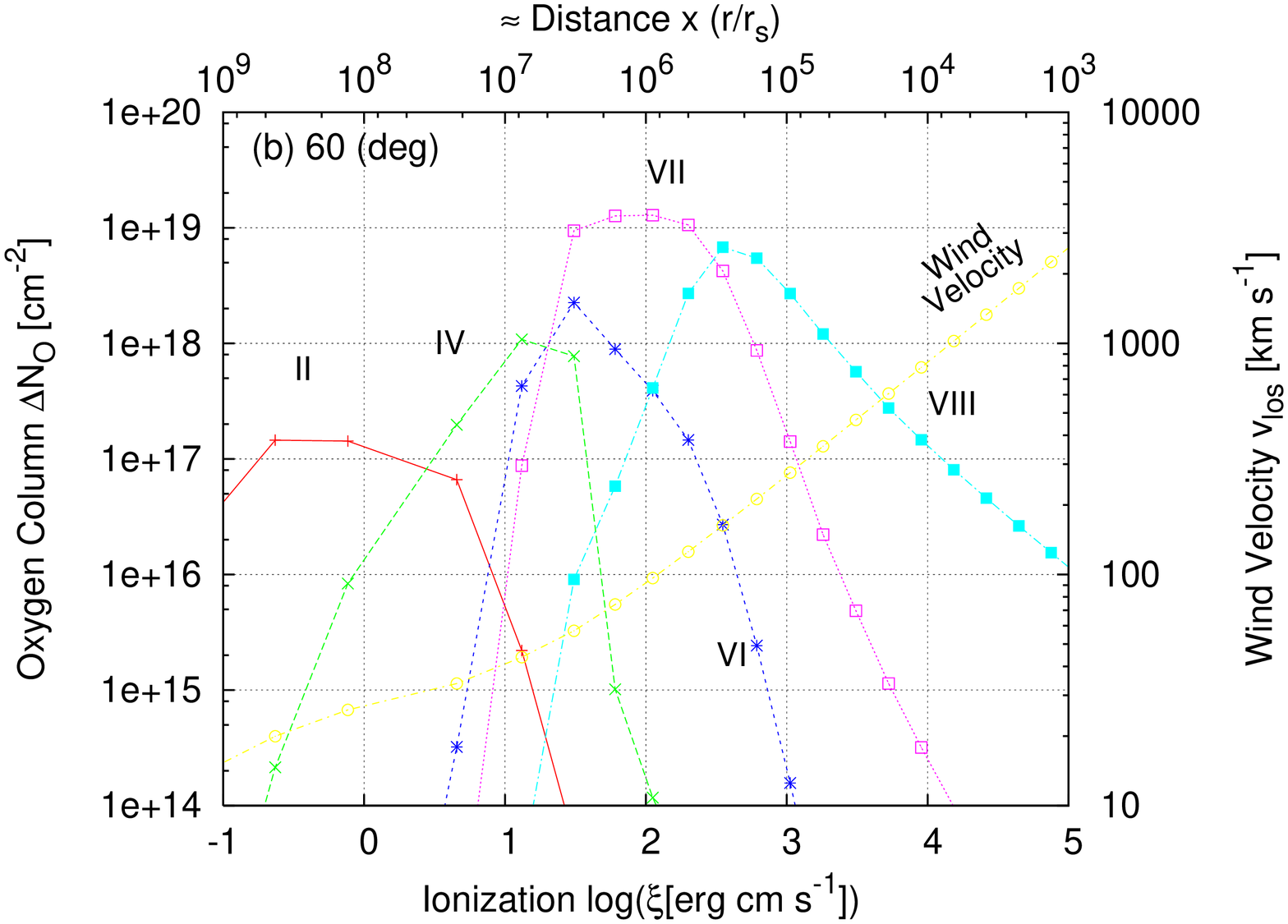,angle=-90,width=0.45\linewidth,clip=""}
\end{tabular}
\caption{Same as Figure~\ref{fig:fe} but for oxygen. [{\it See the
electronic edition of the Journal for a color version of this
figure.}] \label{fig:o}}
\end{figure}

To conclude this section, we show in Figure~\ref{fig:o} the column
densities of specific charge states of oxygen again for two
different LOS at (a) $\theta = 30\degr$ and (b) $\theta=60\degr$,
along with the corresponding velocities overplotted as in
Figure~\ref{fig:fe}. As expected the corresponding outflow
velocities are smaller for \ovii~ ($v \sim 40-150$ km~s$^{-1}$ at
$\log \xi \sim 1-2$ or $\log x \sim 7-8$)  and larger \oviii~ ($v
\sim 100-600$ km~s$^{-1}$ at $\log \xi \sim 6-7$ or $\log x \sim
2-3$) for (a) $\theta = 30\degr$, while similarly for (b) $60\degr$
we find $v \sim 40-150$ km~s$^{-1}$ at $\log \xi \sim 2$ ($\log x
\sim 6$) for \ovii~ and $v \sim 100-300$ km~s$^{-1}$ at $\log \xi
\sim 2.5-3$ ($\log x \sim 5-5.5$) for \oviii, with the former having
lower column per ion than the latter, as expected.

One should note that the precise correlations between the wind
variables and the observables (i.e. $\Delta N_{\rm ion}$, $v_{\rm
los}$) in the present model are primarily a function of the
specified mass-accretion rate $\dot{m}$ and LOS angle $\theta$ as
well as the incident X-ray spectral index $\alpha$ and therefore the
model has the freedom to alter the resulting AMD distributions
depending on the intrinsic properties of AGNs (here we focused on
IRAS~13349+2438 as a case study).

%%---------------------- figure~3 -----------------------------
%\begin{figure}[t]
%\centerline{
%\includegraphics [keepaspectratio=false,width=2.1in,angle=-90]{amd-30deg.eps}
%\includegraphics [keepaspectratio=false,width=2.1in,angle=-90]{fe-30deg.eps}
%\includegraphics [keepaspectratio=false,width=2.1in,angle=-90]{amd-60deg.eps}
%\includegraphics [keepaspectratio=false,width=2.1in,angle=-90]{fe-60deg.eps}
%} \caption{simulated Equivalent $N_H$ distribution for various elements and wind temperature $T$ in the fiducial model. Rapid suppression of $N_H$ at higher $\xi$ arises simply from ignoring fully ionized atoms. }
%\label{fig:rsp1}
%\end{figure}

\subsection{Absorption Line Spectra}

As discussed in the previous subsection and exhibited plainly in
Figures~\ref{fig:fe} and \ref{fig:o}, specific charge states of the
different elements are present at different values of $\xi$ to which
correspond different values of column density and velocity. We have
also indicated that the simplest assumption about the optical depth
of a specific transition is that it is determined  by the velocity
shear of the wind, a quantity that is also calculable within our
models [see equation~(37)]. Armed with this information we present
in this section the profiles of specific atomic transitions in
absorption. With the spectral emissivity and velocity given, we are
also in a position to calculate the corresponding emission profiles.
However, since the most prominent AGN lines lie in the O-UV part of
the spectrum, a comprehensive approach would require also a more
precise model of the UV part of the spectrum; as such we defer line emission to a future work.

\begin{figure}[t]% ------------------------------------- Figure~7
\centering
\begin{tabular}{cc}
\epsfig{file=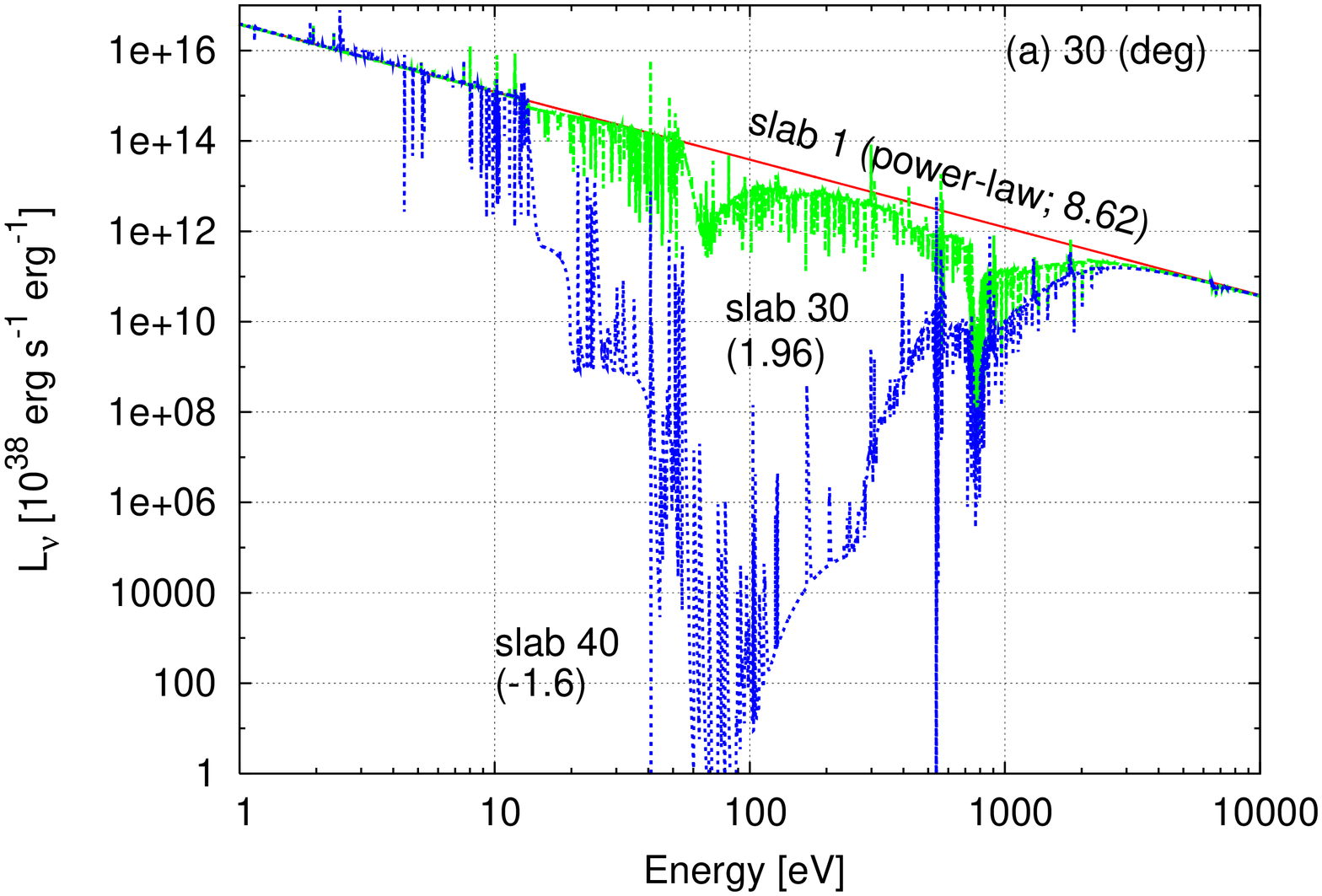,angle=-90,width=0.45\linewidth,clip=""}
&
\epsfig{file=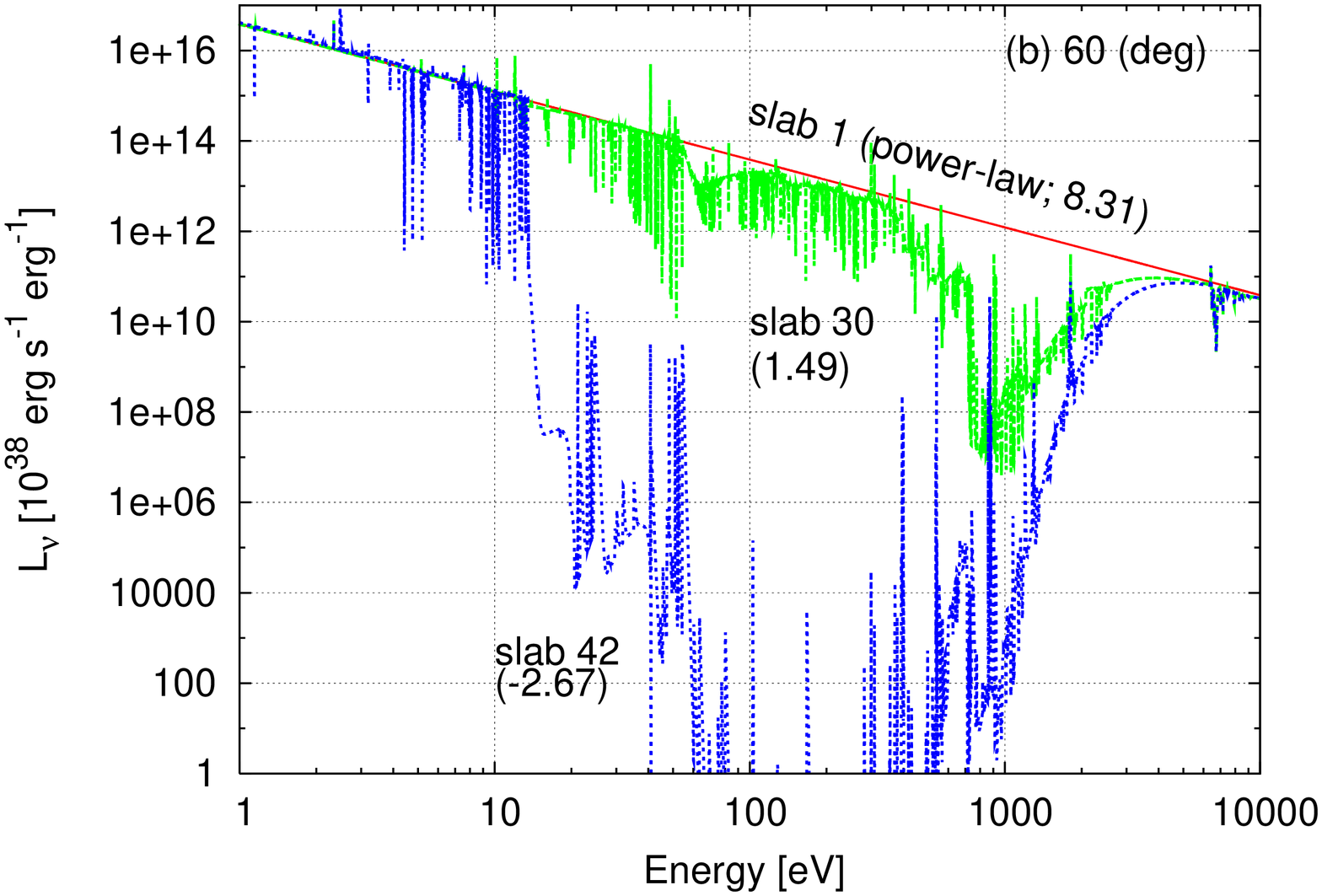,angle=-90,width=0.45\linewidth,clip=""}
\end{tabular}
\caption{Simulated broad-band X-ray spectra $L_{\nu}$ for (a)
$30\degr$ and (b) $60\degr$ as a sequence of progressive X-ray
illumination: an initial power-law (slab 1) and the subsequent
spectra (as indicated by ionization parameter $\log \xi$-value of
each slab) with the same fiducial model parameter as in
Figure~\ref{fig:amd}. [{\it See the electronic edition of the
Journal for a color version of this figure.}] \label{fig:spec}}
\end{figure}

A demonstration of the progressive effect of absorption on the
transmitted spectra is presented in Figure~\ref{fig:spec}. This
figure exhibits the spectra between 1 eV and 10 keV of the radiation
transmitted past the innermost slab 1 (power-law) through the
subsequent slabs indicated by the values of $\log \xi$ for two
different LOS angles, (a) $\theta=30\degr$ and (b) $60\degr$.
Absorption by the consistently ionized wind becomes apparent first
at energies $E \lsim 1$ keV, becoming progressively deeper and
shifting to lower energies as the radiation propagates through
larger wind column. Absorption also increases with the inclination
angle $\theta$ as one would qualitatively expect. Note that in this
simplistic model the incident radiation includes only the X-ray
power-law continuum ignoring all the other (potentially rather
important) components. Also, it ignores emission that eventually may
come into our LOS due to reflection and scattering in the wind. For
this reason, the deep absorption imprinted in Figure~\ref{fig:spec}
should not be interpreted literally in a quantitative sense, but as
a qualitative indicator of the absorption as a function of $\xi$
along a given LOS.

%

%
%-----------------------------------------------Place Figure~8
\begin{figure}[t]% ------------------------------------- Figure~8
\epsscale{1.0} \plottwo{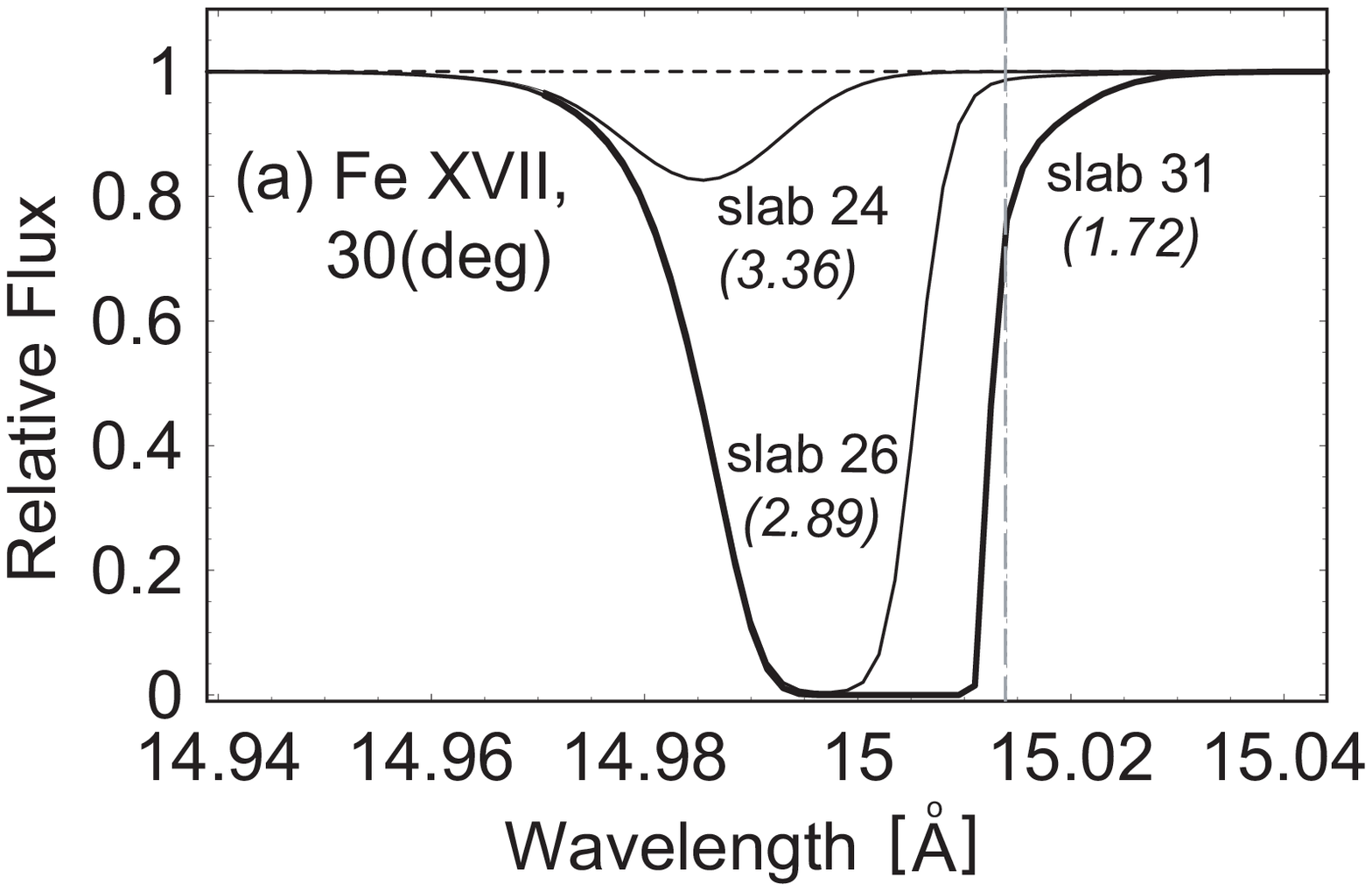}{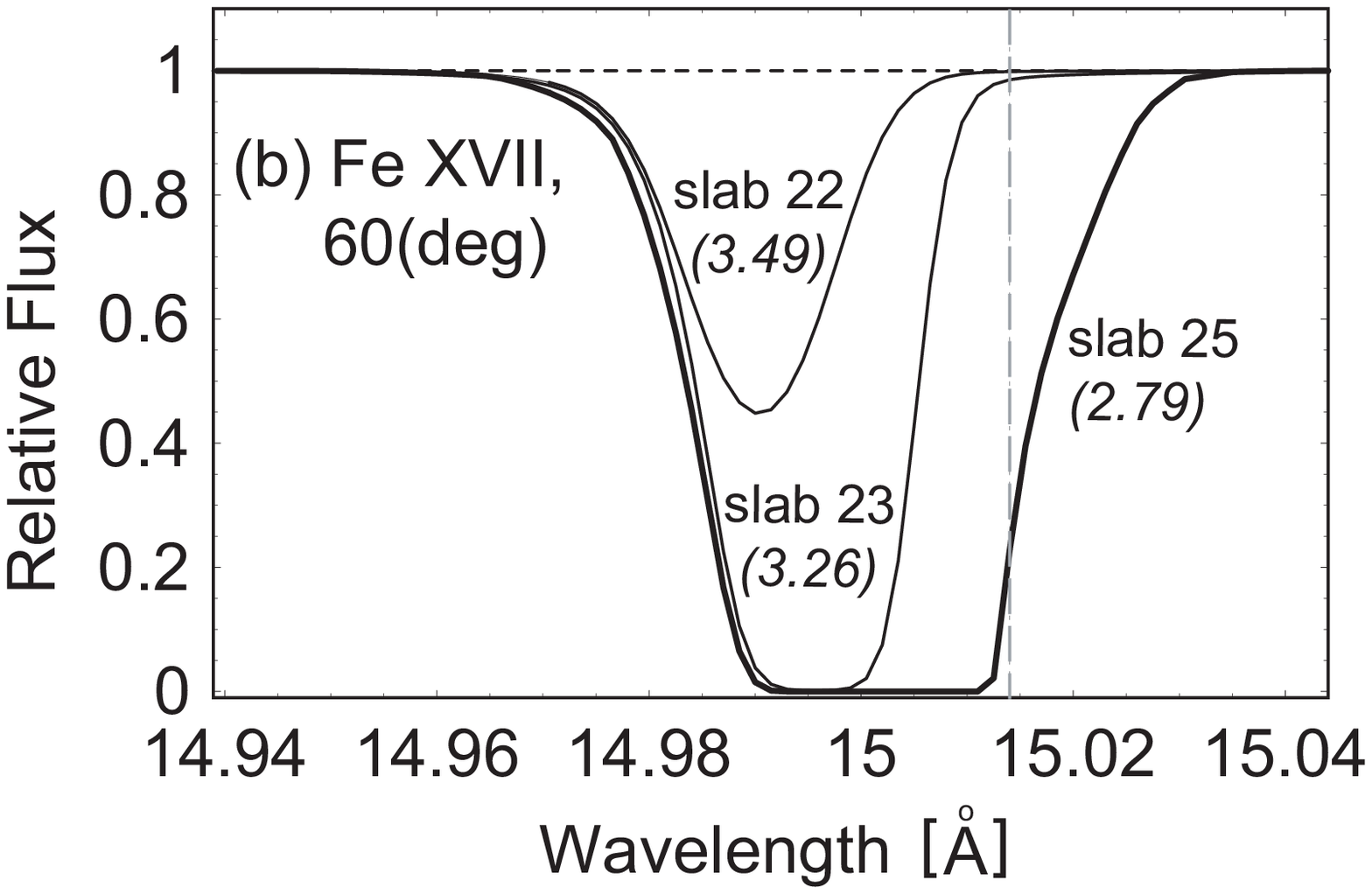}
\plottwo{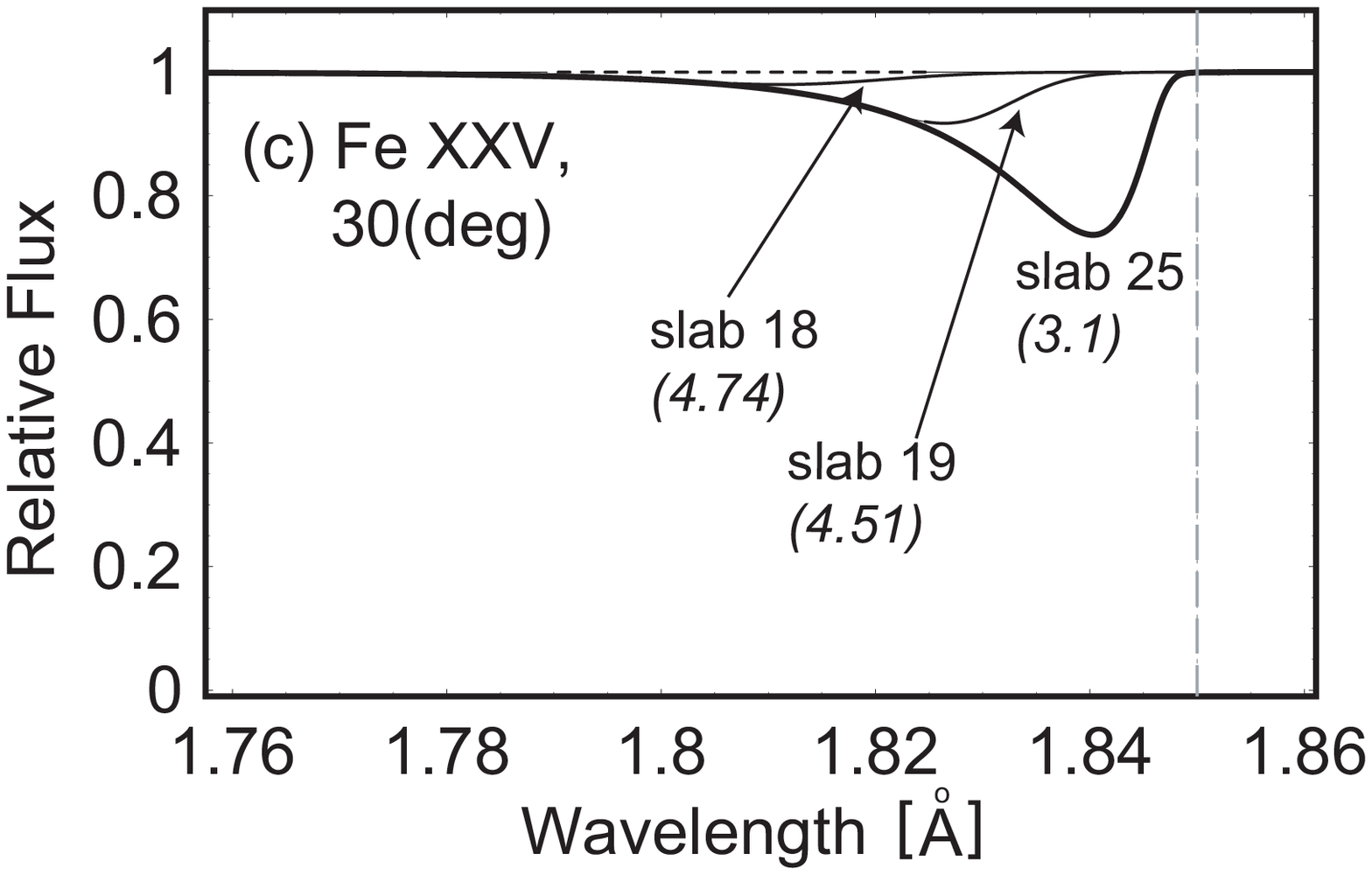}{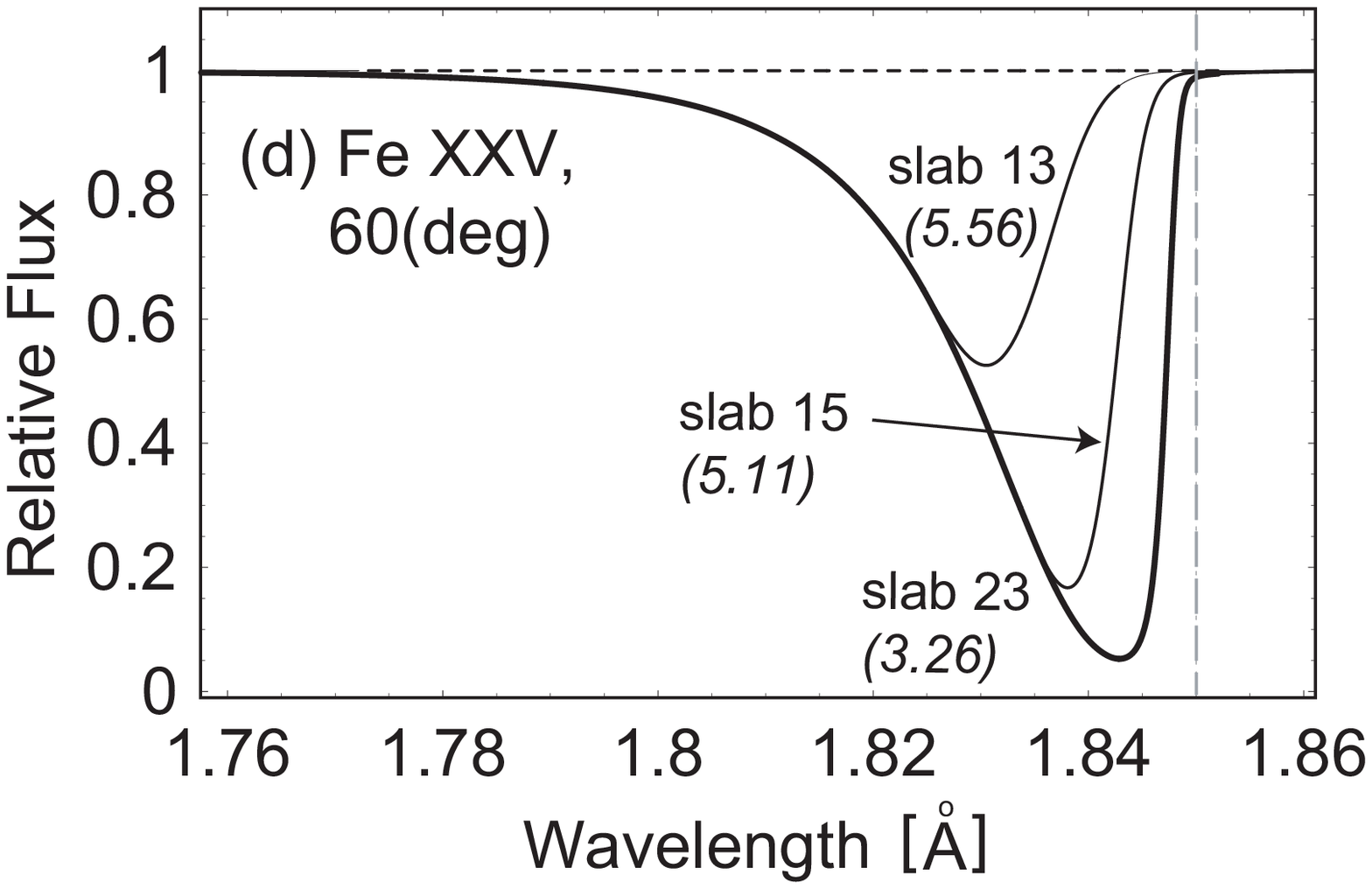} \caption{A sequence of the
simulated absorption features (indicated by ionization parameter
$\log \xi$-value of each slab) for \fexvii~ (upper panels) and
\fexxv~ (lower panels) as the incident X-ray (power-law shown by
dotted horizontal lines) transmits through the LOS wind of $30\degr$
(left panels) and $60\degr$ (right panels) with the same fiducial
model parameters as in Figure~\ref{fig:amd}. Vertical dashed lines
denote the rest-frame wavelength.} \label{fig:abs}
\end{figure}
%-------------------------------------------------------------

Let us turn our attention to specific atomic transitions, namely
those of \fexvii~ and \fexxv~ of column $\Delta N_{\rm Fe}$ with
velocity $v_{\rm los}$ as a function of $\xi$ as given in
Figure~\ref{fig:fe} and shear given by our model. Using the
expressions for the photoionization cross-section and Voigt function
[eqns.~(\ref{eq:tau1})- (\ref{eq:voigt})] one can construct the
characteristic absorption profiles; these include both the Doppler
blueshift due to LOS velocity $v_{\rm los}$ and the line broadening
of the Voigt function in equation~(\ref{eq:voigt}).
Figure~\ref{fig:abs} shows a sequence of the simulated absorption
line for \fexvii~ with the rest-frame wavelength at $15.014${\AA}
(upper panels) and \fexxv~ with the rest-frame wavelength at
$1.850${\AA} (lower panels) for two LOS angles, $\theta=30\degr$
(left panels) and $60\degr$ (right panels). We focus here on these
prominent iron charge states because they are less contaminated by
other adjacent spectral features. We adopted the following values
for the relevant atomic data: $f_{\rm ij}=0.69/2.3$ and $\Gamma=4.48
\times 10^{14}/2.27 \times 10^{13}$ s$^{-1}$ for \fexxv/\fexvii,
respectively. The figure provides the spectra transmitted through
increasing wind column along the LOS. The absorption line obtains
its largest blueshift and broadest contribution by ions at the
higher values of $\xi$, becoming progressively deeper, less
blue-shifted and narrower as the contribution to the absorption is
affected by regions of lower $\xi$ (larger $r$) which provide the
highest column of the specific ion (and also smaller velocities and
shear). Again, the spectra are broader and deeper for the higher LOS
angles reflecting higher (local) velocities at specific values of
$\xi$ and higher overall columns.

In concluding this section we would like to point out the work of
Gabel et al. (2003) on the UV absorption features of NGC 3783 which
were found to have velocity structure in agreement with that of the
corresponding X-ray ones, thereby arguing for the consistency of the
entire wind structure. Also, Collinge et al. (2001) have found that
the low ionization Fe absorption features of NGC 4051 had
corresponding UV absorption features, while the high ionization,
higher velocity X-ray absorption features of the spectrum lacked an
equivalent UV absorption, indicating the absence of these ions in
the higher ionization, higher velocity plasma.

%Again, a progressive pattern of the spectrum given by
%equation~(\ref{eq:tau1}) is determined through the local column and
%its position (i.e. local LOS velocity) obtained in the model AMD
%(e.g. Fig.~\ref{ref:}).

%\subsection{Stability Curves??? }

\section{Summary and Discussion}

In this work we have presented a detailed study of the ionization
structure of model MHD winds off accretion disks; this is our first
attempt to model within this context the recent observations of
absorption features in the {\it ASCA, XMM} and {\it Chandra} X-ray
spectra, the so-called warm absorbers. To this end we have employed
the self-similar 2D hydromagnetic wind models developed by CL94 that
provide the fluid 2D density and velocity fields, which we coupled
to photoionization calculations using \verb"XSTAR". Our
consideration of magnetocentrifugal (MHD) winds/outflows in this
work has been motivated and supported in part by recent
observational implications that at least in a number of AGNs and
GBHCs the inferred driving mechanism of the observed X-ray ionized
wind medium is magnetic rather than thermal or radiative (e.g., see
Miller et al.~2006, 2008 for GRO~J1655-40; Kraemer et al.~2005 and
Crenshaw \& Kraemer~2007 for NGC~4151).
%\citep[][]{Kraemer05,Crenshaw07,Miller08}.

Given the scope of our paper, our model wind is necessarily overly
simple. It ignores a host of issues that affect the structure of
winds off accretion disks and replaces them with the self-similar
models of CL94. The interested reader can get a feeling of the
multitude and complexity of the issues not addressed in the present
treatment by looking at the works of, e.g., \citet[][]{Proga03} and
\citet[][]{ProgKal04} (and references therein) and also \citet[][for disks with several distinct values of the accretion rate $\dot
{m}$]{Ohsuga09}. These include, among others, the radiative transfer in the
wind, the ensuing effects of radiation pressure and also the
influence of the vertical gradient of $B_{\phi}^2$ in the launching
of the wind. As shown in these references the winds can be radiation
driven in the inner part of the disk (due to the increased flux at
this spatial domain) while being magnetically launched at larger
radii. In this respect our model winds are quite different in
structure and variability from the more realistic winds of
\citet{Proga03} at small radii, while they should be more similar to
those of his at larger radii where radiation pressure is smaller and
the wind is magnetically launched. Realistic models must by
necessity consider the transition of the underlying disk from
radiation pressure, at small $r$, to gas pressure dominance at
larger radii; this fact would likely force a different choice of our
boundary condition $\psi'(90^{\circ})$ and therefore yield a field
geometry that breaks the self-similarity of our solutions.  Wind
models very similar to those used herein but with the inclusion of
the effects of the radiation pressure that break the self-similarity
were discussed by KK94 and \citet{Everett05}.

One should therefore view our models with the above caveats in mind.
The significance and assumptions of the models we propose is
basically justified {\em a posteriori} by their ability to interpret
the observations. Self-similarity is one of the fundamental
assumptions of our model winds. However, whether exactly
self-similar or not, any model that would attempt to account the
entire range of ionic species shown e.g. in \citet{Behar09} must by
necessity cover a very broad range in the photoionization parameter
$\xi$. Then, the column of each such ion provides a measure of the
corresponding hydrogen equivalent column $N_H$ as a function of
$\xi$, i.e. the AMD. These two quantities ($N_H, \xi$) can then be
employed to provide a measure of the gas density $n(r)$ as a
function of the distance $r$ from the X-ray source. Barring the
possibility of several independent regions at different distances
but similar columns as indicated by the functional form of the AMD
(one could consider this is possibility at the risk of introducing
an inordinate number of free parameters), models with radial density
profiles as those considered here are a natural consequence of the
AMD obtained by the {\it Chandra} observations.

The wind ionization structure was followed in 1D as described in \S
2.2, namely along the observer's LOS, assuming the ionizing source
to be point-like. Clearly, a more comprehensive treatment of this
problem should take into account also the
backward emitted radiation which will reach the observer from the
regions of the disk on the other side of the black hole, as well as
the scattered radiation, whose effects could be significant. We plan to
return to these issues in a future publication.

The 2D structure of the winds considered in our analysis implies, at
a minimum, a two-parameter description of the absorption features.
However, the self-similarity of the problem simplifies further the
treatment by allowing the separation of the $r$ and $\theta$
variables with the wind density obtaining the form  $n(r, \theta)
\propto r^{2q-3} {\cal N}(\theta)$. The parameter $q$ is a most
important parameter of these models because it determines both the
radial dependence of the ionization parameter $\xi$ and the wind
column density per decade of radius along a given direction (LOS).
In the models examined in the present work we restricted our study
to the value $q=1$, which leads to a radial density profile $n(r)
\propto 1/r$. This is an interesting profile in that it provides for
an ionization parameter $\xi$ with a similar dependence, i.e. $\xi
\propto 1/r$, and most importantly, with equal column per decade in
radius, across the entire range in radius and the corresponding
range in $\xi$; as a result, ionic species of very different
ionization properties, existing over a wide range of ionization
parameter (and radius), have roughly comparable column densities,
independent of the distance and are therefore possible to detect. As
noted in CL94, the value of $q$ determines also the axial current
distribution in radius within the wind, with the corresponding
magnetic energy per unit length at the wind base being also constant
per (cylindrical) decade in radius. It is worth comparing our work
with that of KK94, who used the same type of MHD wind with the value
of $q$ as we do herein: while these authors focused their study on
the effects of dust on radiation transfer, we focus our attention on
the effects of the ionizing radiation on the X-ray spectra of AGNs,
in particular on the ionized absorbers.
%It is worth noting at this point that
KK94 also noted in passing the relevance of their models to the X-ray AGN
spectra well ahead of the detailed outflow observations made with
{\it Chandra}.

While the structure of the wind in $r$ and $\theta$ is provided by
the models of CL94, for the remaining parameters of our models,
namely the normalization of the wind density and the X-ray
luminosity of the ionizing source we have chosen to use the scalings
of ADAFs, or better ADIOS; as a result the X-ray luminosity is
proportional to the square of the (normalized) accretion rate at the
inner edge of the disk $\dot m^2$ rather than simply $\dot m$,
assumed in standard accretion disks.
%One should note that in winds with $q \ne 3/4$ the mass-accretion rate
%depends on the radius, being $\dot m \propto r^{1/2}$ for winds of $q=1$.
Therefore, restricting
ourselves to models with $q=1$, the global ionization structure of a
given wind, including the normalization of the column density,
depends only on {\it two parameters}, namely $\dot m$ and the observer's
inclination angle $\theta$. This parametrization provides an extremely
economical set of assumptions concerning not only the outflows, i.e.
the wind ionization structure seen in absorption in detailed
X-ray spectra, but also for the entire (radio-quiet) AGN
unification picture: Figure \ref{fig:density} which exhibits the
normalization of the wind density as a function of $\theta$ along
the magnetic field line of $\Psi = \Psi_0$, makes
apparent the difference in column between face-on and edge-on views,
implying that the wind, if extending to sufficiently large radii,
can in fact serve as the proposed molecular torus associated with
AGN unification, a point originally made by KK94, also for winds
with $q=1$; it is of interest to note that the few
objects for which sufficiently detailed observations exit are
consistent values $q \simeq 1$ \citep[HBK07;][]{Behar09}. Within this
same context and ionizing luminosity considerations, one should note
the dependence of source obscuration on the X-ray luminosity, namely
its reduced value for objects accreting at a higher fraction of
their Eddington rate $\dot m$, which apparently is in general
agreement with observations. It remains to be seen whether these
notions can withstand the scrutiny of more consistent and
encompassing observational tests.

At this point we would like to stress the importance of the AMD in
the study of AGN outflows/winds, a quantity enunciated by
HBK07 and modeled in detail in the preceding sections.
Analyses similar to those of HBK07 and \citet{Behar09} are
indispensable because they produce a consistent analysis of the
entire set of absorption features in the AGN X-ray absorption
spectra. At the same time they underscore the unique value of X-ray
spectroscopy which, in a wavelength band of $\sim 1.5$ decades, can
encapsulate the properties of ions that span $\sim 5$ orders of
magnitude in ionization parameter and, for models with $q \simeq 1$,
a similar range in radius; at the same time, measurement of their
absorption columns yields the equivalent hydrogen column of the
flow, $N_H$, over a similar range in radius, thereby going a long
way toward the determination of the physics underlying the outflow
dynamics. The fact that the AMD analyses to date are
roughly independent of $\xi$ provide support to our use of the $q=1$
or $n(r) \propto 1/r$ models, reiterating that a similar density
profile was invoked on the basis of AGN and GBHC timing properties
\citep{PNK01,KHT96}.

We have thus presented a concrete example of the AMD dependence on
$\xi$ for our $q=1$ models with $\epsilon = 0.2, ~ \dot m = 0.1$ and
two different values of the observer's inclination angle (see
table~\ref{tab:tbl-1} for details). As expected we found the AMD to
be constant, i.e. independent of $\xi$, over many decades in this
parameter (i.e. the local column density $\Delta N_H$ is independent
of ionization states $\xi$ of ions). With $\dot m = 0.1, ~\eta_W
\simeq 0.5$, equation~(\ref{eq:column1}) implies $\Delta N_H \sim
2.6 \times 10^{21}$ cm$^{-2}$ (yielding a total column of $N_H \sim
3.9 \times 10^{22}$ cm$^{-2}$) for $\theta=30\degr$ and $\Delta N_H
\sim 1.8 \times 10^{22}$ cm$^{-2}$ (total of $N_H \sim 2.5 \times
10^{23}$ cm$^{-2}$) for $60\degr$ over $-1 \lesssim \log \xi
\lesssim 4$ erg~cm~s$^{-1}$, in good agreement with the observed AMD
of IRAS~13349+2438 \citep[HBK07;][]{Behar09} and also some other
AGNs detailed below. One should note that these values depend
primarily on $\dot m$ and $\theta$ and are independent of the black
hole mass. The mass of the object gets involved only as a measure of
its total luminosity, which does not appear in the expression for
$N_H$, implying that these models could in principle be applicable
also in GBHCs. High quality X-ray absorption data are in fact
available for the GBHC GRO~J1655-40 and were used to argue for
magnetic driving of the wind in this system too
\citep{Miller06,Miller08}. These spectra are distinguished from
those of AGNs by the prominent absence of low ionization state ions.
This is to be expected given that the presence of the companion star
limits the extent of the disk to roughly half the distance between
the two objects or $r \simeq 10^{12}$ cm. Given that the
Schwarzschild radius of the compact object is $r_S \simeq 10^6$ cm
the entire disk size covers only a range of $\log x \sim 6$ in
radius; considering (based on Fig.~\ref{fig:amd}) that in the inner
region of the wind ($r \sim 1000-3000 r_S \sim 10^9$ cm) major
elements (even heavier species) are almost fully ionized, one would
expect the presence of ions over only a factor of 1000 in $\xi$, in
rough agreement with the observation.

Given that our models provide also the complete velocity field of
the MHD wind we have also produced a sequence of synthetic
absorption profiles as described in \S 2.2 and shown in \S 3.2. We
have done so for two charge levels of Fe, namely \fexxv~ and
\fexvii. With our fiducial model predicting the \fexvii~ column to
be maximum at $\log \xi \sim 2.2-3$ erg~cm~s$^{-1}$, we then infer
the corresponding LOS velocity (see Fig.~\ref{fig:fe}) to be $v_{\rm
los} \sim 100-300$ km~s$^{-1}$ in excellent agreement with the
observed values of IRAS~13349+2438. It is important to note that the
model produces not only the correct velocity $v_{\rm los}$ at the
maximum value of $\Delta N_{\rm Fe}(\xi)$ but also the observed
$\Delta N_H(\xi)$ normalization {\em for the same inclination
angle}. It is therefore possible with measurements of the combined
line widths and column densities to reproduce within the present
models both $q$, $\theta$ and $\dot m$, thereby providing a complete
specification of these winds. In addition to IRAS~13349+2438, the
nearby bright Seyfert MCG~6-30-15 has been observed to show, as one
of the multiple ionization zones, an X-ray ionized absorber with an
outflow velocity of $\sim 1900$ km~s$^{-1}$ at $\log \xi \sim 3.85$
and $N_H \simeq 9 \times 10^{22}$cm$^{-2}$ \citep[e.g.][]{Young05,
HBA09} also in good agreement with our model results for $\dot m
\sim 0.1$ (see Fig.~\ref{fig:fe}). Clearly, a different choice of a
set of conserved quantities (wind variables) in the model would
produce a slightly different field line geometry in which the
resulting MHD outflow could in principle obtain higher LOS
velocities (perhaps by factors of magnitudes), and this will be
studied in detail in a future work.

As noted above our model is quite successful in reproducing the
observed AMD of IRAS~13349+2438. More recently, \citet{Behar09}
presented a compilation of the AMD of a number of AGN with $d N_H/d
\log \xi$ which are slightly different from constant but with
$q-$values of $n(r) \propto r^{2q-3}$ which are still very close to
unity; e.g. NGC~3783 ($q \sim 0.89$), NGC~5548 ($q \sim 0.94$),
NGC~7469 ($q \sim 0.9$) and MCG-6-30-15 $(q \sim 0.95)$. We believe
some of these AMDs are sufficiently close to those of our fiducial
model to be virtually indistinguishable. By comparison, the MHD wind
models of BP82 have $q=0.75$ and $n(r) \propto r^{-3/2}$. Their
ionization parameter decreases more slowly with $r$ ($\xi \propto
r^{-1/2}$) and hence a range of 5-6 decades in $\xi$ implies a range
of 10-12 decades in radius with the corresponding distances being
unrealistic. Most importantly, the wind columns would decrease with
radius ($N_H \propto r^{-1/2}$) and the AMD dependence on $\xi$
would be $d N_H/d \log \xi \propto \xi$, in clear disagreement with
these observations, and thus their model is essentially ruled out
\citep{Behar09}.

Despite the apparent success of this first model, one should bear
in mind that several aspects have been treated in a rather
simplified fashion. Here we discuss some of them and their
influence on the results presented so far:

1. The shape of the ionizing spectrum used so far ($F_{\nu} \propto
\nu^{-\alpha}$ between 13.6 eV and 13.6 keV with $\alpha = 1.5$) is
quite simplistic, however, such spectra are often used in similar
type calculations \citep[e.g.][]{Sim08}. This is a crude
approximation to the observed spectra characterized by complex
spectral shapes. Spectral features like the BBB and the UV to X-ray
luminosity ratio $\alpha_{\rm OX}$ \citep[][]{Elvis00} play a role
in our results \citep[a recent review of the broad band AGN SEDs can
be found, e.g., in][]{Risaliti04}. More recently, \citet{Grupe04}
sampled 110 bright soft-Xray selected AGNs for a simultaneous study
of Optical/UV and X-ray data and reported strong correlations
between the X-ray spectral slope and the Optical/UV slope which
should be included in a more comprehensive treatment.
Multi-components of the broad-band AGN spectra, as those discussed
by \citet{Elvis00} and \citet{Risaliti04}, should play an important
role in characterizing the observed X-ray spectral features. Adding
softer photons (responsible for the Optical/UV components) could
impact the radiative transfer between those more complex photon
distributions and the ionized matter \citep[see, e.g.,][for
multi-component injected spectra]{Everett05,Sim05}. In fact, we have
mimicked the presence of such softer photons by choosing larger
values of the spectral index of our models ($\alpha \gtrsim 2$). For
the same total luminosity (implemented in the definition of $\xi$),
it was found in this trial run that the radial position of the peak
column density of a given ion (e.g. \fexvii) decreases, because the
presence of a given ion requires a certain flux of {\em ionizing}
photons per atom; in a steep spectrum the proper ratio is found at
larger values of the parameter $\xi$ or correspondingly smaller
values of the distance $x=r/r_S$. As a result, the widths of the
corresponding transitions should be larger and therefore the proper
ionizing spectrum is necessary for the correct interpretation of the
relations between ionic column densities and wind kinematics
\citep[see][for a similar approach]{Everett05,Sim05}.

2. The ionizing source has been considered, for simplicity, to be
point-like. A source of finite size will provide for more complex
illumination, especially for parts of the wind at which the source
angular extent is significant. Most importantly an extended source
would impact the absorption line profiles, since a line of sight may
pass through matter other than that used in section 3.2. Finally, to
treat the radiative transfer correctly at all energies, one should
eventually need to resort to a 2D approach rather than the 1D model
(radial only) presently used, which would modify the global
ionization structure (Fig.~\ref{fig:2d}b).

3. The wind equations, as presently implemented include only MHD
forces, while it is apparent that in the presence of the ionic
species we produce one should also include the effects of
radiation pressure. Because the latter depends on the ionization
state of the plasma which in turn depends on its kinematics,
implemented correctly, this should be done iteratively in a way
similar to \citet{Everett05}. Based on the rather small effect
that the radiation force seems to have on the structure of these
winds, we believe that our present results are generally valid.

In summary, we have presented above a first attempt at interfacing
theoretical models of MHD winds with AGN observations, in particular
the absorption features in their X-ray spectra of ionized outflows.
We have placed most of our emphasis on modeling the corresponding
AMDs, which, based on the existing observations seems to favor a
specific value of the model parameter $q$ which determines the
radial density dependence of our models, namely $q \simeq 1$. With
the value of this parameter set by observation, our models present a
{\it two-parameter} family of AGN wind structure, namely $\dot m$
and $\theta$. We have found that, based on the limited number of
objects discussed above, our models fare rather well in
accommodating the observations and the possibility of incorporating
AGN unification within the same models, as discussed in KK94. There
is still a multitude of issues related to X-ray absorption that
remain open, e.g. whether the model can accommodate the high
velocity ($v/c \sim 0.1 - 0.25$) outflows associated with X-ray
absorption features \citep[see, e.g.,][for some attempts with Monte
Carlo simulations]{Sim05,Sim08} in the spectra of some bright
quasars such as APM~08279+5255, PG~1211+143 and PDS~456
\citep[e.g.][]{Chartas09,Pounds09,Reeves09}. In the context of our
model this point may be well explained with an optimized magnetic
field geometry [i.e. $\Psi(r,\theta)$ distribution determined by
GS-equation~(\ref{eq:GS})] that provides a favorable LOS velocity
component. Important as they are, these go beyond the scope of the
present work and we expect to return to them in future publications.

\acknowledgments

We are grateful to our anonymous referee for his/her constructive comments that improved the manuscript. We would like to thank Tim Kallman for his help with \verb"XSTAR"
incisive comments. We express our gratitude to George Chartas for his comments on the model, Takanori Sakamoto and
Javier Garcia for their assistance with scripting and running
\verb"XSTAR" as well as helpful comments. This work was supported in
part by NASA ADP grant.

\end{document}